\documentclass[10pt, journal, twoside]{IEEEtran}
\usepackage{amsmath}
\usepackage{hhline}
\usepackage{silence}
\usepackage{graphicx}
\usepackage{multirow} 
\usepackage{graphicx}
\usepackage{multirow}
\usepackage{amsmath}
\usepackage[english]{babel}
\usepackage{booktabs}
\usepackage{array}
\usepackage{paralist}
\usepackage{threeparttable}
\usepackage{lipsum}
\usepackage{flushend}
\usepackage{cuted}
\usepackage{algpseudocode}
\usepackage{algorithm}
\usepackage{amssymb}
\usepackage{color}
\usepackage{psfrag}
\usepackage{epsfig}
\usepackage{multicol}
\usepackage{color}
\usepackage{fancybox}
\usepackage{subfigure}
\usepackage{cite}
\usepackage{theorem}
\usepackage{url}
\usepackage[table]{xcolor}
\usepackage{epstopdf}
\usepackage[normalem]{ulem}
\usepackage{siunitx}
\usepackage[justification=centering,font=small,labelfont=bf]{caption}
\definecolor{gray1}{gray}{0.90}
\definecolor{gray2}{gray}{0.98}
\definecolor{light-gray}{gray}{0.95}

\newcommand{\ignore}[1]{}

\begin{document}

\title{A Linear-Time Algorithm for Steady-State Analysis of Electromigration
in General Interconnects}

\author{Mohammad Abdullah Al Shohel, Vidya A. Chhabria, and Sachin S. Sapatnekar\\
\thanks{The authors are with the Department of Electrical and Computer Engineering,
University of Minnesota, Minneapolis, MN 55455, USA.  This research was
supported in part by the NSF under award CCF-1714805, the DARPA OpenROAD
project, and the Louise Dosdall Fellowship, and a University of Minnesota Doctoral
Dissertation Fellowship.
}
}

\maketitle

\begin{abstract}
Electromigration (EM) is a key reliability issue in deeply scaled technology
nodes. Traditional EM methods first filter immortal wires using the Blech
criterion, and then perform EM analysis based on Black's equation on the 
remaining wires.  The Blech criterion is based on finding the steady-state 
stress in a two-terminal wire segment, but most on-chip structures are 
considerably more complex. Current-density-based assessment methodologies, i.e., 
Black’s equation and the Blech criterion, which are predominantly used to detect 
EM-susceptible wires, do not capture the physics of EM, but alternative 
physics-based methods involve the solution of differential equations and are slow. 
This paper uses first principles, based on solving fundamental stress
equations that relate electron wind and back-stress forces to the stress evolution
in an interconnect, and devises a technique that analyzes 
any general tree or mesh interconnect structure to test for immortality.
The resulting solution is extremely computationally efficient and its computation
time is linear in the number of metal segments.  Two variants of the method are
proposed: a current-density-based method that requires traversals of the interconnect
graph, and a voltage-based formulation negates the need for any traversals.
The methods are applied to large interconnect networks for determining the
steady-state stress at all nodes and test all segments of each network for
immortality. The proposed model is applied to a variety of tree and mesh
structures and is demonstrated to be fast. By construction, it is an exact 
solution and it is demonstrated to match much more computationally expensive
numerical simulations.
\end{abstract}

\section{Introduction}
\label{sec:intro}
\noindent
On-chip interconnect wires are plagued by the problem of electromigration (EM).
EM is a reliability failure mechanism that may occur when high currents flow
through wires for long periods, e.g., in supply wires in a digital or analog
circuit.  The underlying mechanism of electron transport that causes these
currents to flow implies that the charge carriers transfer momentum to metal
atoms. Over an extended period, this results in material transport of atoms,
which could eventually lead to the formation of voids, or breaks in the wire,
resulting in open circuits. Although EM has long been a concern in integrated
circuits, the mechanisms have changed significantly in recent years.
EM mass transfer phenomena in copper lines are different from that in
older aluminum lines.  In older technologies, EM was considered a problem only
in upper metal layers that carry the largest current, but in FinFET
technologies, as transistors drive increasing amounts of current through narrow
wires, EM hotspots have emerged as a significant issue in lower metal layers.

The conventional method for EM analysis for interconnects involves a two-stage
process. In the first stage, EM-immune wires are filtered out using the Blech
criterion~\cite{blech:76}, which recognizes that stress in a metal wire can settle
to a steady-state value due to the counterplay between the electric field and
the back-stress. This back-stress is caused by the tendency of electrons to diffuse from regions of
higher concentration, as they mass towards the anode, to regions of lower 
concentration towards the anode. The Blech criterion compares the product of
the current density $j$ through a wire with its length, $l$.   This $jl$
product is compared against a technology-specific threshold, and any wires that
fall below this product are deemed immortal, while others are potentially
mortal.  In the second stage, wires in the latter class undergo further
analysis to check whether or not the EM failure may occur during the product
lifespan.  Traditionally, this involves a comparison of the current density
through these wires against a global limit, set by the semi-empirical Black's
equation~\cite{black:69}; more recent approaches
include~\cite{Chen2016,Chatterjee18,vivek:dac,Mishra16}.

However, the Black/Blech approach is predicated on analyses/characterizations of
single-wire-segment test structures, which determine the critical $jl$ product
threshold for the Blech criterion, and the upper bound on $j$ in Black's
equation. In practice, wires typically have multiple segments that carry
different current densities. The criterion for immortality under this scenario
is quite different from the Blech criterion, and while a few past works have
described these differences, there is no computationally simple test similar to
the Blech criterion to determine immortality for general interconnect
structures.

As opposed to the empirical Black's-equation-based approach, there has been an
emerging thread on using physics-based analysis for EM in interconnects.
Building upon past work such as~\cite{rosen:71,Schatzkes86,Clement92}, the work
in~\cite{kor:93} presented a canonical treatment of EM equations in a metallic
interconnect, with exact solutions for a semi-infinite and finite line.  This
paper has formed the basis of much work since then, with techniques that attempt
to obtain solutions for a single-segment
lines~\cite{Ohring11,Sukharev14,Chen2016}.  For multisegment lines, several
attempts have been made to solve the general transient analysis
problem~\cite{Chen2016,Chen2017,Chatterjee18} through detailed simulations.
However, the key to checking for immortality is to solve the {\em steady-state}
problem, and a few approaches have addressed this problem.  The methods
in~\cite{Riege98,Clement99}, subsequently extended to circuit-level analysis
in~\cite{ala:05,Li15}, used a sum of $jl$ products along wire segments: if
$j_i$ is the current through the $i^{\rm th}$ segment of length $l_i$, then the
largest $\sum j_i l_i$ on any path in a tree was taken to be the worst-case
stress: as observed in~\cite{Abbasinasab15}, this is incorrect. 

An experimentally-driven method in~\cite{parkvianode:10} observed an
apparently counterintuitive observation in multisegment wires: that failures
can occur sooner in segments with lower current density. It presented a
heuristic approach for finding an effective current density. As we will show,
this scenario can be explained using our approach, and a more precise
formulation for the effective current density can be determined.
In~\cite{Haznedar06}, a system of equations describing steady-state analysis in
an interconnect tree was presented and solved. However, the structure of the
difference equations was not exploited to obtain a generalizable solution.  The
analyses in~\cite{LinOates13,Lin16} solve a related problem for a simple two-
or three-segment structure with a passive reservoir.  The work in~\cite{Sun18}
develops analysis principles and applies them to several structures, with
closed-form formulas for simple topologies.  However, it does not provide a
scalable algorithm for general structures, and it runtimes of hours are far larger than our reported runtimes.

Thus, a truly general formula for immortality detection that can replace the Blech
criterion for general structures has not yet been developed.  In this work, we
present a computationally simple framework for detecting immortality within any segment of a general
tree/mesh multisegment interconnect structure.  In particular, we develop theoretical
results that demonstrate the applicability of the method to EM analysis in
general mesh structures and present two techniques for steady-state EM analysis.
For a general interconnect mesh structure with $|V|$ vertices and $|E|$ edges,
the computational complexity of both methods is $O(|V|+|E|)$: this is the first
linear-time method that has been presented for steady-state EM analysis of
general structures.  

The focus of this work is purely on determining an efficient technique to
replace the Blech criterion to filter wires for immortality.  Once this updated
criterion is applied to filter out immortal wires, existing
computationally-expensive
methods~\cite{Chen2016,Chen2017,Chatterjee18,vivek:dac,Mishra16,Li11}, can be
applied to the remaining, potentially mortal, wires to determine whether they
may fail during the chip lifetime. Although recent work~\cite{Shohel21b} points towards directions
for improving the computation time of these methods, the cost of any transient EM
analysis will always be larger than that of a steady-state analysis.

This paper is an extended version of~\cite{Shohel21a}, and has several enhancements.
It is shown that the current-density-based method for computing DC stress, which requires tree
traversals to compute stress, is equivalent to a voltage-based method can compute
stresses without tree traversals. As part of this analysis, we also prove that IR drops
across a wire segment is directly proportional to the difference in stress at its two ends. 
Thus, not only is reducing IR drop consistent with reducing EM susceptibility,\footnote{The
intuition for this is easy to grasp since both require wire widening in power grids.} 
but the relationship is one of direct proportionality.
Additionally, we present an intuitive analysis of the
idea of this paper; a comparison against the via node vector 
method~\cite{parkvianode:10}; and an expanded set of results.

The paper is organized as follows. We present core background about the EM
analysis problem in Section~\ref{sec:background} and then develop our main
results related to steady-state EM analysis in Section~\ref{sec:steady-state}.
We then present two algorithms for fast computations of stress in general interconnect
structures: one using current-density-based tree traversals in
Section~\ref{sec:solution}, and another using voltage-based tree traversals in
Section~\ref{sec:voltage_solution}.  Next, we show a comparison of our approach
with a prior heuristic method in Section~\ref{sec:vianodevector}.  Experimental
results are shown in Section~\ref{sec:results}, followed by concluding remarks
in Section~\ref{sec:conclusion}.

\section{Background}
\label{sec:background}

\begin{figure}
\centering
\includegraphics[width=0.7\linewidth]{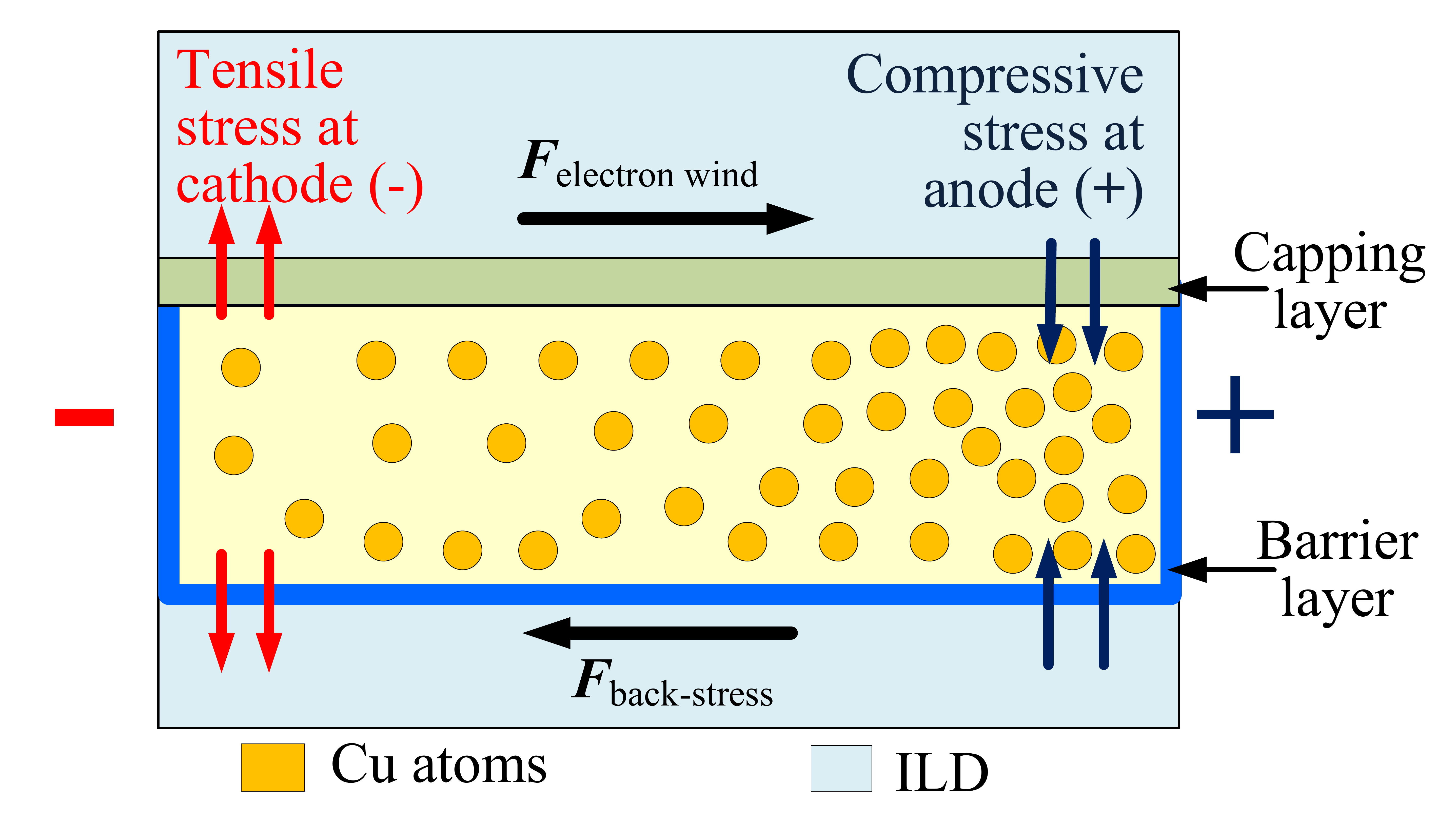}
\caption{Cross section of a Cu wire indicating the electron wind force and
back-stress force~\cite{Mishra16}.}
\label{fig:Cu DD wire}
\end{figure}

\noindent
Figure \ref{fig:Cu DD wire} illustrates the electromigration mechanism in a Cu
dual-damascene (DD) wire.  As the current flows in a metal wire, metal atoms
are transported from the cathode towards the anode, in the direction of
electron flow, by the momentum of the electrons.  This electron wind force
causes a depletion of metal atoms at the cathode, potentially resulting in void
formation, leading to open circuits.  In a Cu DD interconnect, the movement of
migrating atoms occurs in a single metal layer, and atoms are prevented from
migrating to other metal layers due to the capping or barrier layer, which acts
as a blocking boundary for mass transport~\cite{Gambino18,Zhang10}.
Consequently, within a metal layer, mass depletion of atoms occurs at the
cathode terminal and mass accumulation occurs at the anode terminal.  As a
result, a compressive stress is created near the anode, and a tensile stress
near the cathode.  When the tensile stress at the cathode exceeds a threshold,
a void is created.

However, there is another force that acts against this mass transfer.  As metal
atoms migrate towards the anode, the resulting concentration gradient creates a
force that acts against the electron wind force. This back-stress force is
caused by the concentration differential between the anode and cathode sides of
the wire: this creates a diffusion force that is proportional to the stress
gradient. 

\subsection{Notation}

\noindent
For a general interconnect structure with multiple segments, we define the
following notation. This is represented by an undirected graph ${\cal
G}(V,E)$ with $|E|$ segments and $|V|$ nodes.  The vertices $V = \{ v_1,
\cdots, v_{|V|} \}$ are the set of {\em nodes} in the structure, and edges $E =
\{ e_1, \cdots, e_{|E|} \}$ are the set of wire {\em segments}.  A vertex of
degree 1 is referred to as a {\em terminus}.

\begin{figure}
\centering
\includegraphics[width=0.95\linewidth]{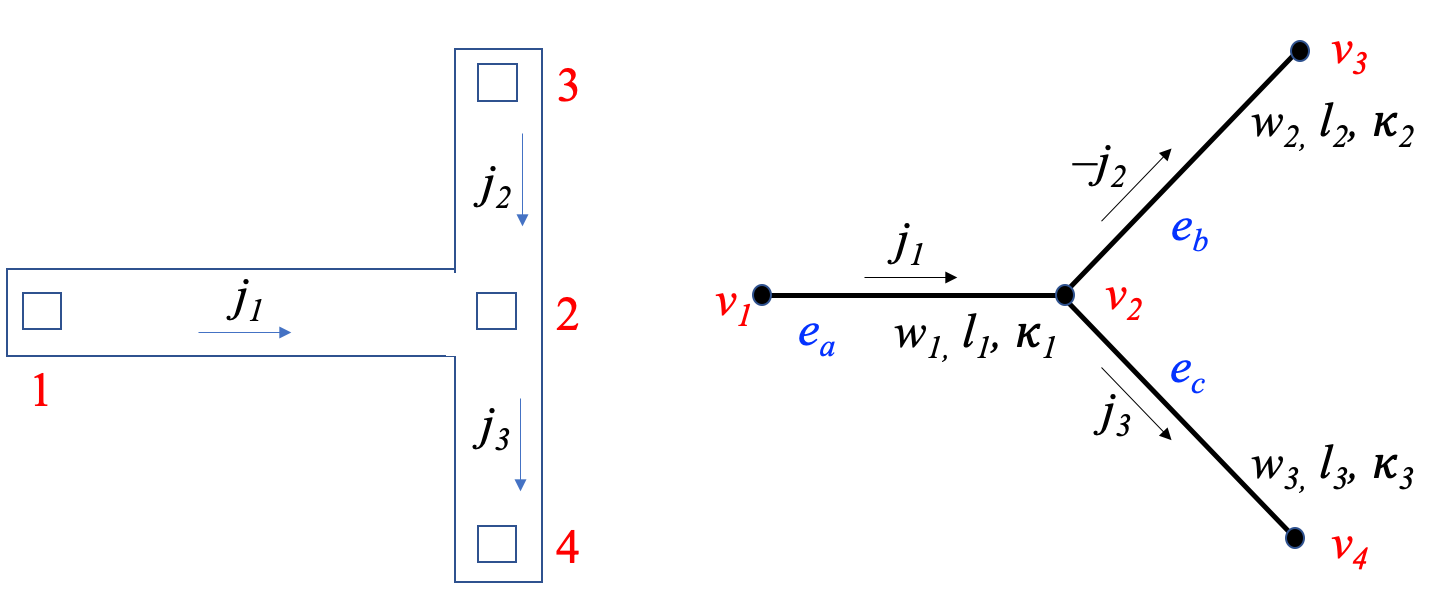}
\caption{(a) A simple net fragment. (b) Its equivalent graph, with arrows showing
the reference current direction for each edge.}
\label{fig:netfragment}
\end{figure}

Each edge $e_i$ is associated with a reference current direction, and has three
attributes: length $l_i$, width $w_i$, and current density $j_i$.  The sign of
current density is relative to the reference direction of the edge: it is
negative if the current is opposite to the reference direction.
Figure~\ref{fig:netfragment} shows a net fragment and its graph model for a
tree with four nodes and three edges: since the current direction in $e_b$ is
opposite to the reference direction, the current density is shown as $-j_2$.

Along each edge, we use a {\em local coordinate system} along each segment $i$.
If the edge has a reference direction from node $v_a$ to node $v_b$, we
represent the position of node $a$ as $x=0$ and that of node $b$ as $x=l_i$.
As part of our analysis, we will compute stresses induced within the
interconnect.  Specifically,
\begin{itemize}
\item $\sigma_i(x,t)$ is the stress within wire {\em segment} $i$ at time
$t$ at a location $x$, where $0 \leq x \leq l_i$ and $1 \leq i \leq |E|$.
\item $\sigma^k$ is the steady-state stress at {\em node} $v_k$,
$1 \leq k \leq |V|$.
\end{itemize}

\subsection{Stress equations for interconnect structures}
\label{sec:stress_structures}

\noindent
A single interconnect segment injects electron current at a cathode at $x=0$
towards an anode at $x=l_i$.  The temporal evolution of EM-induced stress,
$\sigma(x,t)$, in the segment is modeled as~\cite{kor:93}:
\begin{align}
\frac{\partial \sigma}{\partial t} &= 
  \frac{\partial }{\partial x} \left [
         \kappa \left ( \frac{\partial \sigma}{\partial x} + \beta j_i
                \right) \right ]
\label{eq:Korhonen's_eqn}
\end{align}
Here, $x$ is the distance from the cathode and $t$ is the time for which the
wire was stressed and $j_i$ is the current density through the wire.
The term $\beta = (Z^* e \rho)/\Omega$, where $Z^*$ is the effective charge
number, $e$ is the electron charge, $\rho$ is the metal resistivity, and $\Omega$ is
the atomic volume for the metal (in the literature, $\beta j_i$ is often
denoted as $G$).  The symbol $\kappa = D_a {\cal B} \Omega/(kT)$, where $D_a$ is
the diffusion coefficient, ${\cal B}$ is the bulk modulus of the material, $k$
is Boltzmann's constant, and $T$ is the absolute temperature. Further, $D_a=D_0
e^{-E_a/kT}$ where $E_a$ is the activation energy.

When no current is applied, the stress in the wire is given by $\sigma_T$, the
thermally-induced stress due to differentials in the coefficient of thermal
expansion (CTE) in the materials that make up the interconnect stack.  The
differential equation with the boundary conditions can be solved numerically to
obtain the transient behavior of stress over time.  Due to superposition, the
stress in the wire can be computed in this way and $\sigma_T$ can then be added
to account for CTE effects.  The impact of $\sigma_T$ is to offset the critical
stress, $\sigma_{crit}$, to $(\sigma_{crit} - \sigma_T)$.

As in~\cite{kor:93}, the sign convention for $j_i$ is in the direction of
electron current, i.e., opposite to the direction of conventional current and
the electric field.  The atomic flux attributable to the electron wind force is
proportional to the second term on the right hand side that contains $j_i$,
while the flux related to the back-stress force is proportional to the first
term containing the stress gradient $\frac{\partial \sigma}{\partial x}$.  In
both cases, the constant of proportionality varies linearly with the
cross-sectional area of the wire.  The sum, $(\partial \sigma/\partial x +
\beta j_i)$, is proportional to the net atomic flux. 

\noindent
{\bf BCs for single-segment interconnect}
When electron current is injected through the anode and flows to
the cathode at the other end, we have zero-flux conditions at each end:
\begin{align}
\frac{\partial \sigma}{\partial x} + \beta j_1 = 0 \; \;
		\forall \; t \mbox{ at $x=0, x=l_1$.}
\label{eq:BC_single}
\end{align}

\noindent
{\bf BCs for a multisegment interconnect trees/meshes}
The boundary conditions at the terminus nodes (i.e., nodes of degree 1) 
require zero flux across the blocking boundary, i.e.,
\begin{align}
\left . \frac{\partial \sigma_e}{\partial x} 
           \right |_{\mbox{\footnotesize{terminus}}} + \beta j_e  = 0
\label{eq:BC_terminal_tree}
\end{align}
where edge $e$ connected to the terminus has current density $j_e$.

For any internal node $n$ of the structure with degree $d \geq 2$, let
the incident edges with reference current directed into the
node be $\{e_1, \dots , e_m\}$, and the edges directed away from the node be
$\{e_{m+1}, \dots , e_d\}$; if either set is empty, $m=0$ or $d$.  The flux
boundary conditions at such a node are given by
\begin{align}
\sum_{k \in \{1, \cdots , m \}} w_{e_k} 
    &\left ( \left . \frac{\partial \sigma_{e_k}}{\partial x} 
            \right |_n + \beta j_{e_k}
    \right ) = 
\label{eq:BC_internal_tree_flux} \\
& \sum_{k \in \{m+1, \cdots , d \}} w_{e_k} 
    \left ( \left . \frac{\partial \sigma_{e_k}}{\partial x} 
            \right |_n + \beta j_{e_k}
    \right )
\nonumber
\end{align}
and the continuity boundary conditions are:
\begin{align}
& \sigma_{e_1} |_n = \sigma_{e_2} |_n = \cdots = \sigma_{e_d} |_n
\label{eq:BC_continuity_tree} 
\end{align}
where $\sigma_{e_k} |_n$ and $\partial \sigma_{e_k}/\partial x |_n$ are the
values of the stress and its derivative at the location corresponding to node
$n$.

\section{Intuitive Explanation}
\label{sec:intuition}

\noindent
Explanations of electromigration are often mired in the solutions of the
partial differential equations presented in the previous section. In this
section, we present an intuitive view of the solution of the steady-state
problem, which helps easier understanding of the material in the succeeding
sections.

\begin{figure}
\centering
\includegraphics[width=0.7\linewidth]{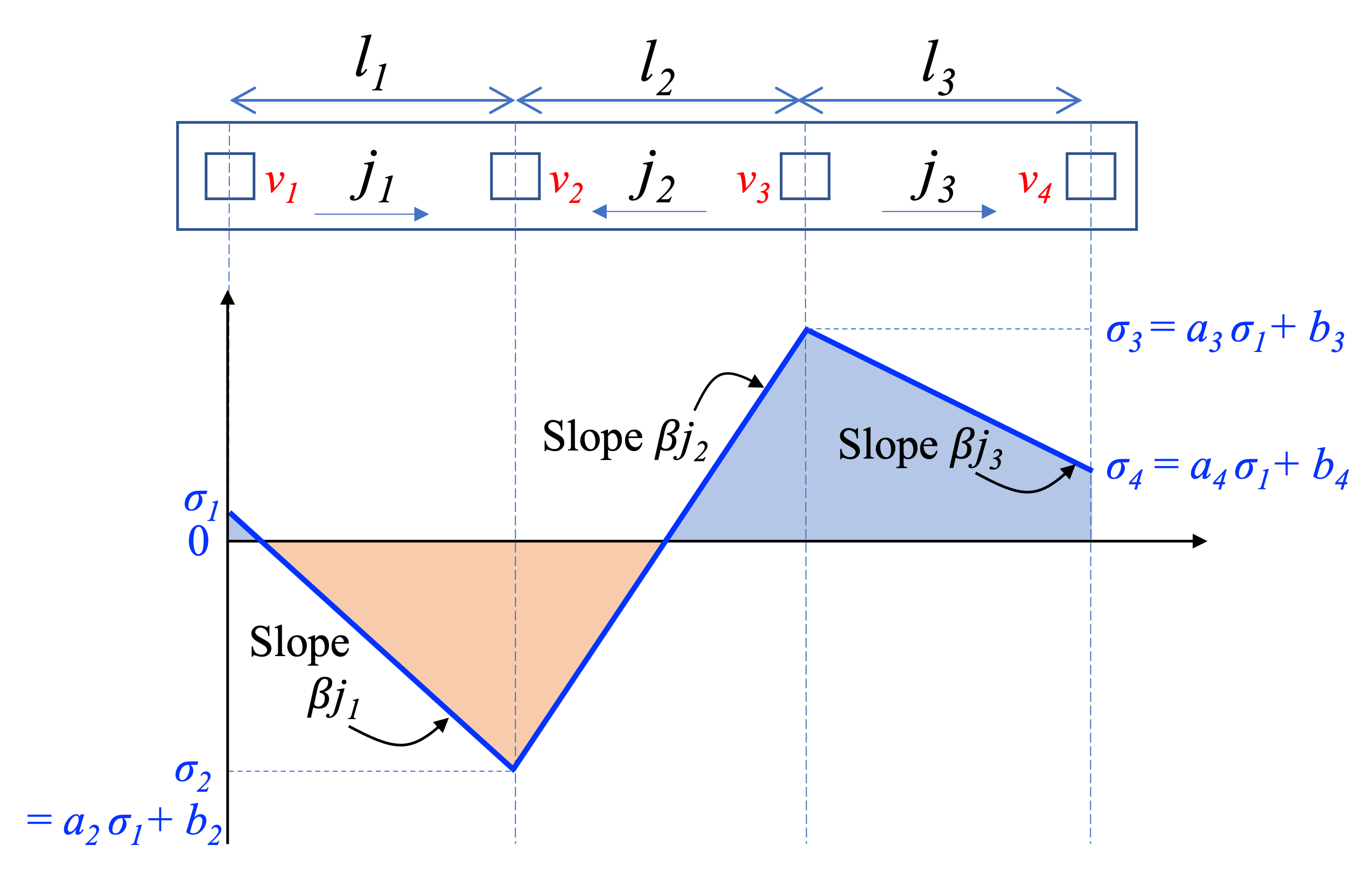}
\caption{A three-segment example to illustrate the intuition behind fast
steady-state computation.}
\label{fig:3seg_example}
\end{figure}

It is well known that in the steady state, after all transients have dissipated,
the stress varies linearly along a wire, and the gradient of the stress is
proportional to the current density in a wire (we will show in
Eq.~\eqref{eq:constant_slope} that for a wire with current density $j$, this
slope is $\beta j$), and that the stress function is continuous at segment
boundaries (Eq.~\ref{eq:BC_continuity_tree}).  Given these two facts, the
stress profile for a three-segment line is as shown in
Fig.~\ref{fig:3seg_example}, where the steady-state stress at vertex $v_i$ is
$\sigma_i$, for $i = 1, \cdots, 4$, and the slope of each segment is
proportional to the current density in the segment; note that the sign of the
slope in the middle segment is the opposite to that in the other segments since
the direction of current flow is also the opposite to other segments.

Therefore, if we knew $\sigma_1$, the other stress values could be computed based
on the known slopes: $\sigma_2$ is lower than $\sigma_1$ by $\beta j_1 l_1$;
$\sigma_3$ is higher than $\sigma_2$ (due to the change in current direction)
by $\beta j_2 l_2$; and $\sigma_4$ is reduced from $\sigma_3$ by $\beta j_3
l_3$.  In other words, given all $j_i$ values, all stresses may be written in
terms of $\sigma_1$ as shown in the figure (as we will show later, this is
equally true for tree structures and meshes), where the values of $a_2, a_3,
a_4, b_2, b_3, b_4$ depend on the currents in the lines.

To find $\sigma_1$, we simply appeal to the concept of conservation of mass: 
although atoms are transported along the wire, there is no change in the total
number of atoms in the wire due to EM. For wires of equal width and thickness,
as in this example, this translates to a condition (Lemma~3) that the integral
of stress over the region is zero. Since all $\sigma_i$ values are expressed
in terms of $\sigma_1$, this integral depends purely on $\sigma_1$: equating it
to zero provides $\sigma_1$, and therefore, $\sigma_2, \cdots, \sigma_4$.

We develop this intuition to a formal set of results in this paper.  We present
two solutions for stress computation:
\begin{itemize}
\item
A {\em current-desnity-based} approach that traverses all segments -- in this case,
from $v_1$ to $v_4$ -- to find the $j_i l_i$ ``drop'' along each segment $i$ to
compute the stress value at each vertex. Combining this with the mass
conservation equation, this provides the steady-state stress at each node.
\item
A {\em voltage-based} formulation where stress computation is traversal-free -- by showing that since
the $j_i l_i$ ``drop'' along a segment is proportional to the IR drop along 
the segment, we point out that the traversal has already been implicitly performed during circuit simulation,
which computes all voltages (and currents) in the system.  Therefore, since
(modified) nodal simulation of the network is essential to predict the currents
through the wires, we show that the steady-state stress at each node can be
computed by eschewing any traversal during EM analysis, and instead, reusing the
steady-state voltages.
\end{itemize}
Both solutions have a cost that is linear in the number of segments in the
circuit, though the latter method has a lower constant of proportionality for
the linear-time complexity since it can be performed without traversals.

We also prove that the EM analysis for mesh-based structures merely needs the analysis of a spanning tree of the mesh, thus making steady-state EM analysis of meshes simple.
\section{Analysis of Steady-state Stress}
\label{sec:steady-state}

\subsection{Equations for steady-state analysis in a wire segment}
\label{sec:segment_solution}

\noindent
We will work with \eqref{eq:Korhonen's_eqn} as a general representation
of the stress in any multisegment line or tree.  In the steady state, when the
electron wind and back-stress forces reach equilibrium, then for each segment
$i$, over its entire length, $0 \leq x \leq l_i$,
\begin{align}
& \frac{\partial \sigma_i}{\partial x} + \beta j_i  = 0,
\mbox{  i.e.,  } \frac{\partial \sigma_i}{\partial x} = -\beta j_i
\label{eq:constant_slope}
\end{align}
The Blech criterion for immortality in a single-segment line asserts that in the
steady state, if the maximum stress falls below the critical stress,
$\sigma_{crit}$, required to nucleate a void, then the wire is considered
immortal, i.e., immune to EM.  This translates to the condition~\cite{blech:76}:
\begin{align}
j l \leq  (jl)_{crit} 
\label{eqn:Blech_criterion}
\end{align}
where $(jl)_{crit}$ is a function of the critical stress, $\sigma_{crit}$. 

The derivation of the Blech criterion is predicated on the presence of blocking
boundary conditions at either end of a segment carrying constant current, and
is invalid for multisegment wires, even though it has been (mis)used in that
context.  For a general multisegment structure, from \eqref{eq:constant_slope},
{\em a linear gradient exists along each segment} of a general multisegment
structure (this has been observed for multi-segment
lines~\cite{Riege98,Clement99} and meshes~\cite{Haznedar06}).  

\noindent
{\em Lemma~1}: 
For edge $e_k$ with reference current direction from vertex $v_a$ to
$v_b$, the steady-state stress along the segment is:
\begin{align}
\sigma_k(x) &= \sigma^a -\beta j_k x 
\label{eq:linearstress1} \\
\mbox{ and }
\sigma^b - \sigma^a &= -\beta j_k l_k
\label{eq:linearstress2}
\end{align}
where $\sigma^a$ ($\sigma^b$) denotes the steady-state stress at node $a$ ($b$).

\noindent
{\em Proof:}
The first expression follows directly from~\eqref{eq:constant_slope}, and the
second is obtained by substituting $x=l_k$ at node $v_b$.
\hfill $\Box$

The following corollary follows directly from~\eqref{eq:linearstress1}:\\
{\em Corollary 1}:
For edge $e_k = (v_a,v_b)$ in an interconnect structure, 
\begin{align}
\int_0^{l_k} \sigma_k(x) dx =
\int_0^{l_k} (\sigma^a - \beta j_k x) dx =
\sigma^a l_k - \beta j_k \frac{l_k^2}{2}
\label{eq:segment_integral}
\end{align}

\noindent
{\em Corollary 2}: In a segment, the largest stress is at an end point.

\noindent
{\em Proof:}
This follows from~\eqref{eq:linearstress2}: if $j_k \geq 0$, the stress on the
segment is maximized at node $v_a$; otherwise at node $v_b$.
\hfill $\Box$

\subsection{Equations for steady-state analysis in a general structure}
\label{sec:general_solution}

\noindent
The existence of cycles in a graph requires careful consideration: we show
that the solution can be found by analyzing a spanning tree.

\begin{figure}[htb]
\centering
\includegraphics[width=0.33\linewidth]{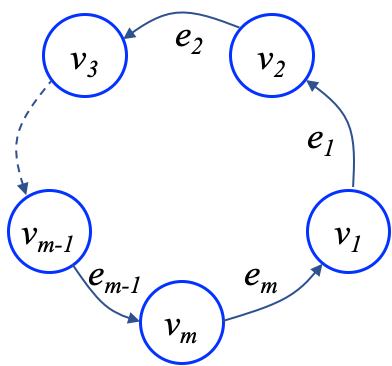}
\caption{A cycle in ${\cal G}(V,E)$.}
\label{fig:cycle}
\end{figure}

\noindent
{\em Theorem 1:} 
Consider any undirected simple cycle, without repeated vertices or edges, $\cal
C$ in ${\cal G}(V,E)$, consisting of edges $e_1, \cdots, e_m$ containing
vertices $v_1, v_2, \cdots, v_m$, with edge reference directions from $v_i$ to
$v_{i+1}$ (where $v_{m+1} \stackrel{\Delta}{=} v_1$), as shown in
Fig.~\ref{fig:cycle}.  The $m$ steady-state stress
equations~\eqref{eq:linearstress2} representing this cycle
are linearly dependent.  A linearly independent set of equations
is obtained by dropping one equation, i.e., breaking the cycle by dropping
one edge.

\noindent
{\em Proof:}
Let $V_i$ be the voltage at vertex $v_i$, $R_i$ be the resistance of wire
segment $i$, $\rho$ be the wire resistivity, and $h_i$ be the wire thickness
(constant in layer $i$). Then $R_i = \rho l_i/(w_i h_i)$ and by Ohm's
law, the electron current $j_i$ is given by:
\begin{align}
j_i = (V_{i+1} - V_{i})/(R_i w_i h_i) = 
(V_{i+1} - V_{i})/(\rho l_i)
\end{align}
According to~\eqref{eq:linearstress2}, along each edge $e_i = (v_i, v_{i+1})$,
\begin{align}
\sigma^{i+1} - \sigma^i &= -\beta j_i l_i = -\beta (V_{i+1} - V_i)/\rho
\label{eq:linearstress3}
\end{align}
Adding up all equations~\eqref{eq:linearstress3} around the cycle, the left
hand side sums up to zero, because each $\sigma^k$ term in one equation has a
corresponding $-\sigma^k$ term in the next equation (modulo $m$, so that
$-\sigma^1$ and $\sigma^1$ appear in the last and first equation,
respectively).  Similarly, the right-hand side also sums up to zero due to
telescopic cancelations of $V^k$ in each equation and $-V^k$ in the next
equation (modulo $m$).

Therefore, the $m$ equations~\eqref{eq:linearstress3} are linearly dependent.
They can be represented by $m-1$ equations: by breaking the cycle at an
arbitrary position and removing one edge, the simple cycle is transformed to a
path with a set of independent linear equations.
\hfill $\Box$

The implications of Theorem~1 are profound, namely: \\
{\em The steady-state stress in any structure with cycles can be solved by
removing edges to make it acyclic, yielding a spanning tree structure, which is
then solved to obtain the stress at all nodes.}

\subsection{Solving the steady-state analysis equations}
\label{sec:tree_solution}

\noindent
We will first analyze a tree structure, since, as shown above, the steady
state difference equations~\eqref{eq:linearstress2} are to be solved over a
spanning tree of a general interconnect structure.  

We choose an arbitrary leaf node of the tree as a reference; without loss of
generality, we will refer to it as node $v_1$, and the stress at that node as
$\sigma^1$.  For any node $v_i$ in the tree, there is a unique directed path
${\cal P}_i$ from $v_1$ to $v_i$, where each edge $e_k = (v_{s,k},v_{t,k}) \in
{\cal P}_i$ has a direction from $v_{s,k}$ to $v_{t,k}$ where $v_{s,k}$ is the
vertex that is closer to $v_1$.  Note that edges on this path are {\em
directed} from $v_1$ towards $v_i$.  However, it is built on an {\em
undirected} graph for the tree, where each undirected edge of the tree
has a reference current direction.

\begin{figure}
\centering
\includegraphics[width=0.7\linewidth]{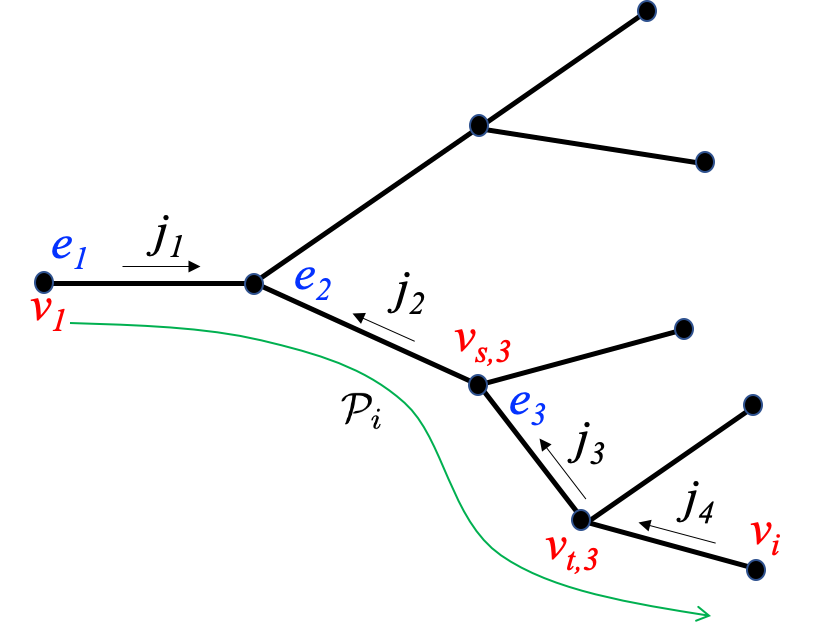}
\caption{An example undirected graph of a tree-structured interconnect, 
showing path ${\cal P}_i$ from reference node $v_1$ to node $v_i$.}
\label{fig:tree_ex}
\end{figure}

To illustrate this point, consider the tree in Fig.~\ref{fig:tree_ex}, with 
path ${\cal P}_i$ from vertex $v_1$ to $v_i$. Vertex $v_{s,3}$ is the vertex of
$e_3$ that is closer to $v_1$.  The reference current directions on the
undirected graph are as shown: the direction of $j_1$ is along the direction of
path ${\cal P}_i$, while $j_2$, $j_3$, and $j_4$ are in the opposite direction.

\noindent
{\em Definition}:
We define $B_{{\cal P}_i}$, the ``Blech sum'' for a path ${\cal P}_i$, as:
\begin{equation}
B_{{\cal P}_i} = \sum_{e_k \in {\cal P}_i} \hat{j}_k l_k
\label{eq:Bi}
\end{equation}
where the summation is carried out over all edges $e_k$ on path ${\cal P}_i$.
The term $\hat{j}_k = j_k$ if the reference current direction for edge $e_k$
is in the same as path ${\cal P}_i$; otherwise, $\hat{j}_k = -j_k$.
Informally, $B_{{\cal P}_i}$ is the algebraic $(j l)$ sum along ${\cal P}_i$
from $v_1$ to $v_i$.

In the example of Fig.~\ref{fig:tree_ex}, the Blech sum to $v_{s,3}$ is
\begin{equation*}
B_{{\cal P}_{s,3}} = j_1 l_1 - j_2 l_2
\end{equation*}

\noindent
{\em Lemma~2}:
The stress, $\sigma^{i}$ at node $v_i$ is related to $\sigma^1$ as follows:
\begin{align}
\sigma^i = \sigma^1 - \beta B_{{\cal P}_i} \label{eq:sigma_i}
\end{align}

\noindent
{\em Proof:}
In a tree, the path ${\cal P}_i$ must be unique~\cite{Cormen09}.  Along this
path, the current on each edge $e_k$ from $v_{s,k}$ to $v_{t,k}$ is $\hat{j}_k$,
i.e., $j_k$ if the reference current direction is from $v_{s,k}$ to $v_{t,k}$,
and $-j_k$ otherwise.  Therefore, from~\eqref{eq:linearstress2}, 
\begin{eqnarray}
\sigma^{t,k} - \sigma^{s,k} = -\beta \hat{j}_k l_k
\label{eq:diffconstraints}
\end{eqnarray}
The continuity boundary condition~\eqref{eq:BC_continuity_tree} ensures that
the stress at the distal end of an edge on ${\cal P}_i$ is identical to that on
the proximal end of its succeeding edge, i.e., for successive edges $e_k$ and
$e_l$ on ${\cal P}_i$, $\sigma^{t,k} = \sigma^{s,l}$.  Therefore, adding these
equations over all edges on path ${\cal P}_k$, we see that as successive edges
on the path share a vertex $v$, $\sigma^v$ cancels out telescopically, except
for $v = v_1$ or $v_i$.  Meanwhile, the $\beta \hat{j}_k l_k$ terms add up, so
that the sum of all equations yields
\begin{align}
\sigma^i - \sigma^1 = - \beta 
               \sum_{e_k \in {\cal P}_i} \hat{j}_k l_k
\end{align}
This leads to the result in~\eqref{eq:sigma_i}. \hfill $\Box$

However,~\eqref{eq:sigma_i} in Lemma~2 stops short of determining $\sigma^i$ at
each node: for a tree with $|V|$ nodes, the lemma provides ($|V|-1$) linear
equations in $|V|$ variables, leading to an underdetermined system where each
node stress is related to the stress, $\sigma^1$, at an arbitrarily chosen leaf
node, $n_1$.  The $|V|^{\rm th}$ equation is obtained from the principle of
the conservation of mass: atoms are transported along a wire, but with zero net
change in the number of atoms in the wire.

\noindent
{\em Lemma~3}:
For a general tree/mesh interconnect with $|E|$ edges, with edge $k$
having width $w_k$ and height $h_k$,
\begin{align}
\sum_{k=1}^{|E|} w_k h_k \int_0^{l_k} \sigma_k(x) dx = 0
\label{eq:masscons}
\end{align}

\noindent
{\em Proof:}
The stress on a wire segment causes a displacement of $u_i$ in segment $i$ of
the interconnect structure.  The stress has no shear component since the
current in a line is unidirectional.  Due to conservation of mass, the net
material coming from all $|E|$ wire segments is zero, and therefore,
\begin{align}
\textstyle \sum_{k=1}^{|E|} w_k h_k u_k = 0
\label{eq:massconv}
\end{align}
where $w_k$ is the width of the $k^{\rm th}$ wire segment.  The displacement
$u_k$ is the integral of displacements $du_k$ over the segment caused by
stress $\sigma_k(x)$ applied on elements of size $dx$ in segment $k$.  If
${\cal B}$ is the bulk modulus, from Hooke's law,
\begin{align}
u_k = \textstyle \int_0^{l_k} du_k(x) 
    = {\cal B} \textstyle \int_0^{l_k} \sigma_k(x) dx
\end{align}
Combining this with~\eqref{eq:massconv} leads to the result of Lemma~3.
\hfill $\Box$
In effect, this result is an integral form of the
BCs~\eqref{eq:BC_internal_tree_flux}, which conserve flux at the boundary of
each segment in the tree.  

\noindent
{\em Theorem 2}:
A tree or mesh interconnect with $|E|$ edges and $|V|$ vertices is immortal
when:
\begin{align}
\max_{1 \leq i \leq |V|} & \left ( \sigma^i \right ) < \sigma_{crit}
     \label{eq:max_vs_crit} \\
\mbox{where  }
\sigma^i &= 
\beta \left [ \frac{\sum_{k=1}^{|E|} w_k h_k
	\left [ \hat{j}_k \frac{l_k^2}{2} - B_{{\cal P}_{s,k}} l_k \right ]}
     {\sum_{k=1}^{|E|} w_k h_k l_k} - B_{{\cal P}_i} \right ]
     \label{eq:sigmaifinal}
\end{align}
where $B_{{\cal P}_i}$ is the ``Blech sum'' defined in~\eqref{eq:Bi}.

\noindent
{\em Proof:}
We first show that expression~\eqref{eq:sigmaifinal} provides the stress at
node $n_i$ of the interconnect, and is obtained by combining the result of
Lemma~3 with the $(|V|-1)$ equations from~\eqref{eq:sigma_i}.

Let edge $e_k$ connect vertices $v_{s,k}$ and $v_{t,k}$, where $v_{s,k}$ is the
vertex that is closer in the tree to the reference node $v_1$. Then,
substituting the result of Lemma~2 into Corollary~1,
\begin{align}
\int_0^{l_k} \sigma(x) dx =
\left ( \sigma^1 - \beta B_{{\cal P}_{s,k}} \right ) l_k
	- \beta \hat{j}_k \frac{l_k^2}{2}
\end{align}
where $B_{{\cal P}_{s,k}}$ is the Blech sum from node $n_1$ to node
$v_{s,k}$.\footnote{The use of $\hat{j}_k$ allows for the traversal
from $v_1$ to $v_i$ to include edges in a direction opposite to the reference
current direction: the stress difference between nodes on such edges should
have the opposite sign as~\eqref{eq:linearstress2} in Lemma~1.}

Substituting the integral expressions in~\eqref{eq:masscons} from Lemma~3:
\begin{align}
\sum_{k=1}^{|E|} w_k h_k \left [
\left ( \sigma^1 - \beta B_{{\cal P}_{s,k}} \right ) l_k
	- \beta \hat{j}_k \frac{l_k^2}{2} \right ] = 0
\label{eq:masscons2}
\end{align}
After further algebraic manipulations, we obtain
\begin{align}
\sigma^1 =
  \frac{\beta \sum_{k=1}^{|E|} w_k h_k
	\left [ \hat{j}_k \frac{l_k^2}{2} + B_{{\cal P}_{s,k}} l_k \right ]}
     {\sum_{k=1}^{|E|} w_k h_k l_k}
\label{eq:sigma1}
\end{align}
Finally, we substitute the above into~\eqref{eq:sigma_i} to
obtain~\eqref{eq:sigmaifinal}, the expression for the steady-state stress
values at each node $i$.

For the interconnect to be immortal, the largest value of stress in the tree
must be lower than $\sigma_{crit}$, the critical stress required to induce a
void.  From Corollary 2, in finding the maximum stress in the tree, it is
sufficient to examine the stress at the nodes of the tree, so that the largest
node stress is below $\sigma_{crit}$. This proves~\eqref{eq:max_vs_crit}.
\hfill $\Box$

\section{Linear-Time Immortality Calculation Based on a Current Density Formulation}
\label{sec:solution}

\noindent
As we have established, a general interconnect on a graph can be solved by
considering the solution of Theorem~1 on a tree of the graph.  Identifying such
tree is straightforward, and standard methods such as depth-first or
breadth-first traversal can be used.

After arriving at a tree structure, although Theorem~2 provides a useful,
closed-form result, a simple-minded computation would calculate $\sigma^i$ at
each node $v_i$ in the tree through repeated incantations
of~\eqref{eq:sigmaifinal}.  However, as we will show, this computation can be
performed in $O(|E|)$ time for a structure with $|E|$ edges.  We
rewrite~\eqref{eq:sigmaifinal} as:
\begin{align}
\sigma^i &= \beta \left [ \frac{Q}{A} - B_{{\cal P}_i} \right ]
\label{eq:sigmai_rewritten} \\
\mbox{where   } 
Q &= \textstyle \sum_{k=1}^{|E|} w_k h_k \left [
           \hat{j}_k \frac{l_k^2}{2} + B_{{\cal P}_{s,k}} l_k \right ] 
\label{eq:Qcomp} \\
A &= \textstyle \sum_{k=1}^{|E|} w_k h_k l_k 
\label{eq:Acomp}
\end{align}
This computation requires the calculation of three summations for 
$A$, $Q$, and for the Blech sum, $B_{{\cal P}_i}$ from reference node $v_1$ to
each node $i$ in the tree.  It proceeds in the following steps:
\begin{enumerate}
\item[1.]
To compute $B_{{\cal P}_i}$, we traverse the tree from $v_1$ using a standard
traversal method, e.g., the breadth-first search (BFS). At node $v_1$, we
initialize $B_{{\cal P}_{v_1}} = 0$. As we traverse each edge $e_k =
(v_{s,k},v_{t,k})$, we compute $B_{{\cal P}_{t,k}}$.  
\item[2.]
Using the above Blech sums to each node, we compute $Q$ (Eq.~\eqref{eq:Qcomp})
and $A$ (Eq.~\eqref{eq:Acomp}), summing over all edges.
\item[3.]
Finally, we compute $\sigma^i$ at each node $i$
using~\eqref{eq:sigmai_rewritten}.
\end{enumerate}
If the stress at each end of a segment is below $\sigma_{crit}$, it is immortal. If not, it is potentially mortal and is further analyzed using a transient stress analysis method~\cite{Chen2016,Chen2017,Chatterjee18,vivek:dac,Mishra16,Li11,Shohel21b} to determine its stress value at the end of the chip lifetime.

\noindent
{\bf Complexity analysis}: The BFS traversal in Step~1 over a tree traverses
$O(|E|)$ edges. For each edge, Step~2 performs a constant number of
computations to obtain $A$ and $Q$ (~\eqref{eq:Acomp}--\eqref{eq:Qcomp}). The
final computation of~\eqref{eq:sigmai_rewritten} in Step~3, and the immortality
check that compares the computed value with $(\sigma_{crit}-\sigma_T)$
according to~\eqref{eq:max_vs_crit}, perform a constant number of computations
for $|V|$ nodes.  Therefore, the computational complexity is $O(|V|+|E|)$.

\begin{figure}[htb]
\centering
\includegraphics[width=0.5\linewidth]{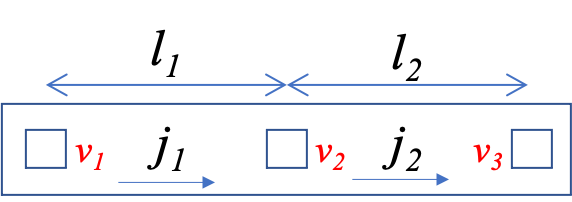}
\caption{A two-segment interconnect line.}
\label{fig:twoseg}
\end{figure}

\noindent
{\bf Example}:
We illustrate our computation for a two-segment line (Fig.~\ref{fig:twoseg}) in
a single layer (with constant $h_k$) in Table~\ref{tbl:twoseg}, using the
leftmost node $v_1$ as the reference.  Starting from $v_1$, the two edges are
traversed to compute $B$.  The symbol $B_{{\cal P}_{t,k}}$ represents the Blech
sum calculated at the distal vertex $v_{t,k}$ of the edge; note that the
computation of $Q$ uses the Blech sum at the proximal vertex, $v_{s,k}$.

\begin{table}[hbtp]
\centering
\caption{Sequence of computations for a two-segment wire.}
\begin{tabular}{|l|c|c|c|}
\hline
                 & $A$                 & $B_{{\cal P}_{t,k}}$ & $Q$ \\ \hline \hline
Initialization   & 0                   & 0                    & 0 \\ \hline
Edge $(v_1,v_2)$ & $w_1 l_1$           & $j_1 l_1$            & $w_1 j_1 l_1^2/2$
\\ \hline
\multirow{2}{*}{Edge $(v_2,v_3)$}
                 & \multirow{2}{*}{$w_1 l_1 + w_2 l_2$}
                                       & \multirow{2}{*}{$j_1 l_1 + j_2 l_2$}
                                                              & $w_1 j_1 l_1^2/2 + 
                                                                  w_2 j_2 l_2^2/2$ 
\\
                 &                     &                      &  $+ w_2 l_2 (j_1 l_1)$
\\ \hline
\end{tabular}
\label{tbl:twoseg}
\end{table}

Based on the table, we compute the stress at each node as:
\begin{align}
\sigma^{v_1} = \beta \frac{w_1 j_1 l_1^2 + w_2 j_2 l_2^2 + 2 w_2 j_1 l_1 l_2}
                          {2(w_1 l_1 + w_2 l_2)} \label{eq:ss_line_stress} \\
\sigma^{v_2} = \sigma^{v_1} - \beta (j_1 l_1)
\; \; ; \; \;
\sigma^{v_3} = \sigma^{v_1} - \beta (j_1 l_1 + j_2 l_2) \nonumber 
\end{align}
The analysis of this line in~\cite{Sun18} yields an identical result;
unlike our method, \cite{Sun18} cannot analyze arbitrary trees/meshes in
linear time.

\section{An Alternative Voltage-Based Formulation}
\label{sec:voltage_solution}

\noindent
In a typical design flow, an interconnect system is analyzed via circuit
simulation to determine currents and voltages throughout the system. For
example, for a power delivery network, which is the most critical on-chip
interconnect system that requires EM analysis, a system of nodal or modified
nodal equations $G \mathbf V = \mathbf J$ is solved to obtain voltages
throughout the network, and currents are then inferred from the (modified)
nodal formulation.  In this section, we will show how an immortality check can
be performed using the computed nodal voltages, without the need for any
traversals during stress computation.

We begin by introducing a result that relates the $jl$ drop along a segment
to the voltage drop across it.\footnote{This result can also be a powerful tool 
in formulating power grid optimization problems. To ensure that EM constraints 
in a grid are met, it is sufficient to ensure that the steady-state EM stress 
(which is a sum of $jl$ terms) does not exceed
$\sigma_{crit}$. Given this direct relationship between EM and IR drop constraints,
it is possible to formulate a power grid optimization problem that ensures EM-safe
power grids while working {\em entirely} within the space of voltage variables.}

\noindent
{\em Lemma 4}: Consider a wire connecting nodes $s$ and $t$, with resistance
$R_{st}$, resistivity $\rho$, and length $l_{st}$.  If the current density in the wire is $j_{st}$, in the direction of electron current, then 
\begin{equation}
j_{st}l_{st} = (V_t - V_s)/\rho
\label{eq:jlV}
\end{equation}
where $V_s$ and $V_t$ are the voltages at endpoints $s$ and $t$ of the wire.
Note that this mirrors the second inequality in Eq.~\eqref{eq:linearstress3}.

\noindent
{\em Proof}: If the conventional current in the wire is $I_{st}$ and its
cross-sectional area is $A_{st}$, then
\begin{align}
j_{st}l_{st} = -(I_{st}/A_{st}) l_{st} = -I_{st} R_{st} / \rho = (V_t - V_s)/\rho
\label{eq:jL_IR}
\end{align}
The second equality arises from the relation $R_{st} = \rho l_{st}/A_{st}$. The
negative sign arises because $j_{st}$ is in the direction of electron current
while $I_{st}$ is in the direction of conventional current.
\hfill $\Box$

\noindent
{\em Corollary 3}: Consider a path ${\cal P}_i$ with edges $e_1, e_2, \cdots, e_k$
and nodes $v_1, v_2, \cdots, v_{k+1}$ with node voltages $V_1, V_2, \cdots, V_{k+1}$,
respectively.  The Blech sum along this path is 
\begin{equation}
B_{{\cal P}_i} = (V_{k+1} - V_1)/\rho
\label{eq:Bi_voltage}
\end{equation}
The proof of this corollary is trivial, and arises through the application of
Eqs.~\eqref{eq:Bi} and \eqref{eq:jlV}: along a path, the voltages of
all intermediate nodes telescopically cancel, leaving the first and last nodes
of the summation as shown in the result.

\noindent
{\em Theorem 3}:
Given the voltages at each node in any interconnect system (tree or mesh),
the EM-induced stress at node $i$ in any interconnect system (tree or mesh), is
given by
\begin{align}
\sigma^i
= \frac{\beta}{\rho} \left [ \frac{\sum_{k=1}^{|E|} w_k h_k l_k V_{av,k}}{A} - V_i \right ] 
\label{eq:sigmai_rewritten_voltage}
\end{align}
where $V_i$ is the voltage at node $i$ and $V_{av,k} = (V_{t,k} + V_{s,k})/2$
is the average of the voltages at the two terminals of the segment represented
by $e_k$.

\noindent
{\em Proof}:
We rewrite $Q$ from \eqref{eq:Qcomp} using voltage variables.  For an
arbitrarily chosen vertex $v_1$ whose node voltage is $V_1$, and which has a
path to the vertex represented by node $i$,
\begin{align}
Q &= \sum_{k=1}^{|E|} w_k h_k l_k \left [
           \frac{\hat{j}_k l_k}{2} + B_{{\cal P}_{s,k}} \right ] \nonumber \\
  &= \sum_{k=1}^{|E|} w_k h_k l_k \left [
           \frac{V_{t,k} - V_{s,k}}{2 \rho} + \frac{V_{s,k} - V_1}{\rho} \right ] \nonumber 
\end{align}
\begin{align}
  &= \sum_{k=1}^{|E|} w_k h_k l_k \left [
           \frac{V_{t,k} + V_{s,k}}{2 \rho} - \frac{V_1}{\rho} \right ] \nonumber \\
  &= \sum_{k=1}^{|E|} w_k h_k l_k \left [
	        \left .  V_{av,k} / \rho - V_1  \right / \rho \; \right ] \nonumber \\
  &= \sum_{k=1}^{|E|} w_k h_k l_k \left [ \left .  V_{av,k} \right / \rho \;             \right ] - A V_1 / \rho
\label{eq:Qcomp_voltage} 
\end{align}
where $A$ is defined in Eq.~\eqref{eq:Acomp}.  From
Eqs.~\eqref{eq:sigmai_rewritten},~\eqref{eq:Bi_voltage}, and~\eqref{eq:Qcomp_voltage},
\begin{align}
\sigma^i
&= \beta \left [ \frac{Q}{A} - \frac{V_i - V_1}{\rho} \right ] \nonumber \\
&= \beta \left [ \frac{\sum_{k=1}^{|E|} w_k h_k l_k V_{av,k}}{A \rho}
- \frac{V_1}{\rho} - \frac{V_i - V_1}{\rho} \right ] \nonumber \\
&= \frac{\beta}{\rho} \left [ \frac{\sum_{k=1}^{|E|} w_k h_k l_k V_{av,k}}{A} 
- V_i \right ] \nonumber 
\hspace{3.2cm} \Box
\end{align}

This leads to the following traversal-free DC stress computation procedure:
\begin{enumerate}
\item [1.]
For each edge $k$ in the circuit, we compute $V_{av,k}$ as the mean of the 
voltages at its two ends in $O(|E|)$ time.
\item [2.]
Over all edges in the circuit, we compute
$\sum_{k=1}^{|E|} w_k h_k l_k V_{av,k}$ in $O(|E|)$ time.
\item [3.]
Next, we compute $A$ using Eq.~\eqref{eq:Acomp}, summing over all edges, also in $O(|E|)$ time.
\item [4.]
Finally, we compute $\sigma^i$ at each node $i$
using~\eqref{eq:sigmai_rewritten_voltage}, given $V_i$, in $O(|V|)$ time.
\end{enumerate}
{\bf Complexity analysis}: Clearly, the cost of the computation is $O(|V|+|E|)$,
including final $O(|V|)$ immortality check that compares the computed value
with $(\sigma_{crit}-\sigma_T)$, requires $O(|V|)$ time. 

It is easily verified that our approach results in the same solution for the
example in Fig.~\ref{fig:twoseg} as listed in Table~\ref{tbl:twoseg}.

A similar voltage-based result has been derived in~\cite{Demircan14} (the precise equations have
some differences but can be shown to be equivalent), but
its potential for linear-time computation was never realized as it was applied
to a set of very small test structures.  This
formulation was used again in~\cite{Sun18}, and applied to a wider set of structures,
however, their results show a quadratic to cubic growth in runtime with problem size,
and it was apparently not noticed that such an approach can yield a linear-time
solution for any arbitrary interconnect, as is shown in this paper. Moreover,
although~\cite{Sun18} applied the technique to mesh structures, neither their proof
nor that in~\cite{Demircan14}, on which the work was based, theoretically
demonstrates that this extension to meshes is valid. In constrast, our work
proves the validity of this method on mesh structures through Theorem~1.

\section{Comparison with the \\Via Node Vector Method}
\label{sec:vianodevector}

\noindent
In~\cite{parkvianode:10}, the two-segment structure of Fig.~\ref{fig:twoseg} was
experimentally analyzed, with a current density of $j_1 = j$ and $j_2 = 2j$,
$w_1 = w_2 = w$, and $l_1 = l_2 = l$. It was observed that the time-to-failure
of the segment with the lower current density was shorter, apparently
contradicting the conventionally held belief that segments with the highest
current are the most vulnerable to EM.
The discrepancy was explained by correctly stating that the underlying cause is
that Segment~1 provides atomic flux to Segment~2, leading to higher depletion at
its cathode. 

\begin{figure}
\centering
\includegraphics[width=0.6\linewidth]{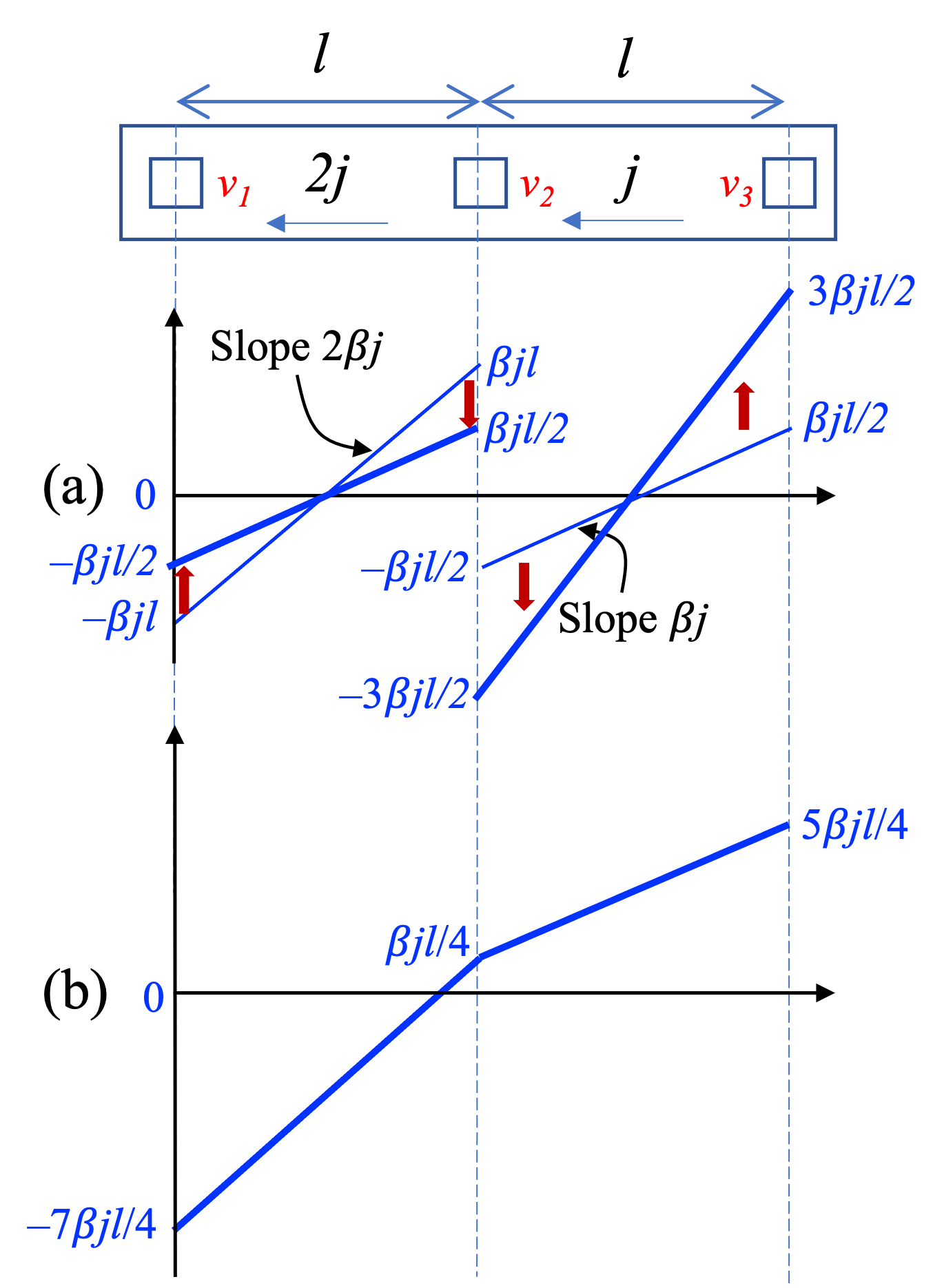}
\caption{A comparison of the results of (a) the via node vector method with 
(b) an exact physics-based analysis for a two-segment line with equal segment
lengths, carrying a current of $2j$ in the left segment and $j$ in the right 
segment.}
\label{fig:vianodevector}
\end{figure}

However, the difference in the time-to-failure between the two
segments was corrected only heuristically by stating that for flux purposes,
Segment~1 has an ``effective'' current of $j + 2j$, corresponding to its own 
flux and the flux supplied to Segment~2. The ``effective'' current on Segment~2
is conjectured to be $2j - j$, accounting for the current of $j$ supplied by
Segment~1.  Based on this, and the observed superlinear behavior of
electromigration time-to-failure (TTF), it was conjectured that a 3$\times$
factor between the effective currents of $3j$ and $j$ on the two segments,
which leads to a 3$\times$ stress differential, could explain the difference in
the time to failure.  However, in a Black's equation framework, this would lead
to an exponent of $>2$ to explain the $\sim$10$\times$ discrepancy in the TTFs.

If the proposed effective currents are applied to the segments independently,
the corresponding stress waveforms are as shown in
Fig.~\ref{fig:vianodevector}(a).  This is clearly incorrect since it does not
satisfy the basic requirements of stress continuity at vertex $v_2$.
A more exact analysis of the steady-state flux can be performed applying
Equation~\eqref{eq:ss_line_stress}, with $j_1 = -2j$ and $j_2 = -j$, to compute 
the peak steady-state stress in Segment~1 and Segment~2 as:

\begin{align}
\sigma^{v_1} &= \beta \left [ \frac{(3 j_1 + j_2)}{4} \right ] l 
= -\frac{7}{4} \beta j l
                          \label{eq:ss_line_stress_accuarate} \\
\sigma^{v_2} &= \beta \left [ \frac{(3 j_1 + j_2)}{4} \right ] l
                          - j_1 l
= \frac{1}{4} \beta j l \nonumber \\
\sigma^{v_3} &= \beta \left [ \frac{(3 j_1 + j_2)}{4} \right ] l 
                          - j_1 l - j_2 l
= \frac{5}{4} \beta j l
\nonumber 
\end{align}

The corresponding stress waveforms, which are correct according to the 
underlying physics,
are shown in Fig.~\ref{fig:vianodevector}(b).  The peak stress on
Segment~1 is actually 5$\times$ that of Segment~2.  Translating this back into
an ``effective'' current, this implies that the effective current is 5$\times$
larger in Segment~1.  For a $\sim$10$\times$ difference in TTF, this leads to a
more conventional exponent for $j$ in Black's equation of 1.4, lying between
the generally accepted range of 1 and 2.

\begin{figure*}
\resizebox{1.05\linewidth}{!}{
\centering
\hspace*{-8mm}
\includegraphics[width=0.33\linewidth]{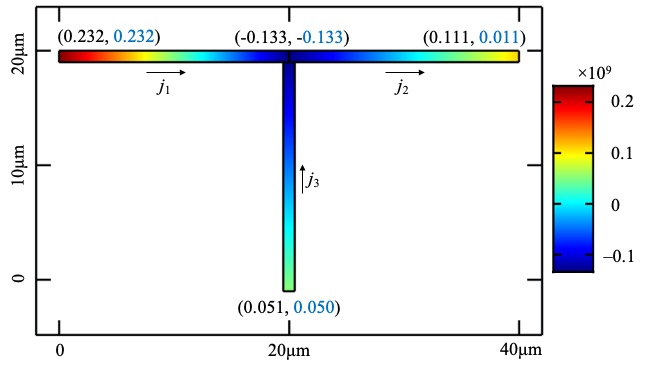}
\hspace*{-3mm}
\includegraphics[width=0.33\linewidth]{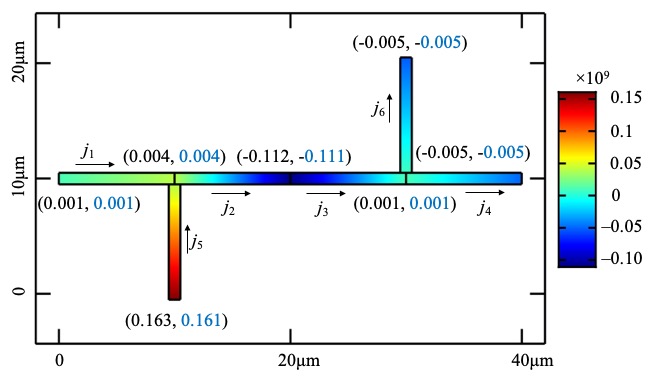}
\hspace*{-3mm}
\includegraphics[width=0.31\linewidth]{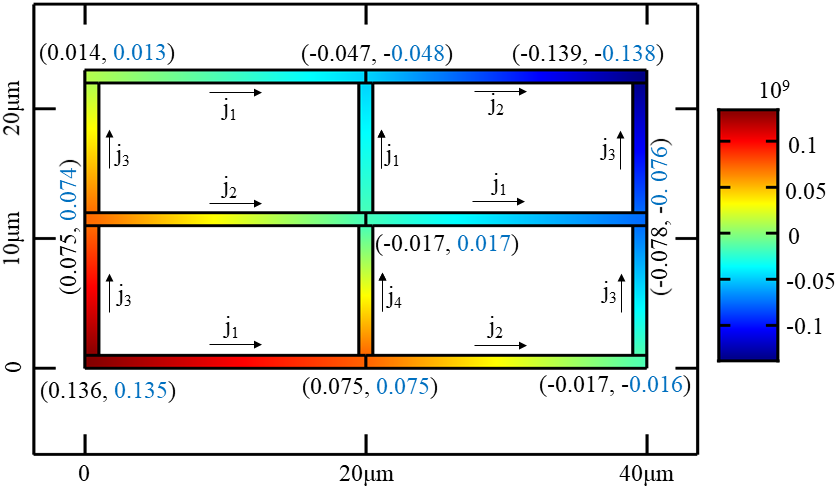}
}
\caption{Comparison of the steady-state stress in three structures: a T,
a tree, and a mesh.  The stress at each node (in GPa) is shown as a tuple, with
our closed-form solution in black and the COMSOL solution in blue text.  The
color bar is based on a COMSOL solution.  The width of each segment
is $1\mu$m, and length scales are shown in the figure. The current densities
in the T are $j_1  = 6 \times 10^{10} A/m^2 $ , $j_2 =
-4  \times 10^{10}$A/m$^2$ , $j_3 = 3\times 10^{10}$A/m$^2$. 
For the tree, $j_1 = -1 \times 10^{10}$A/m$^2$, $j_2 = 5 \times 10^{10}$A/m$^2$,
$j_3 = -4\times 10^{10}$A/m$^2$, $j_4 = j_6 = 2 \times
10^{10}$A/m$^2$, $j_5 = 4 \times 10^{10}$A/m$^2$.  For the mesh structure, $j_1 =
1\times 10^{10}$A/m$^2$, $j_2 = 1.5\times 10^{10}$A/m$^2$, $j_3 =  2\times 10^{10}$A/m$^2$,
$j_4 = 3\times 10^{10}$A/m$^2$.}

\label{fig:COMSOL_color_map}
\end{figure*}

\section{Results}
\label{sec:results}

\noindent
We present three sets of results. We first compare our approach with a numerical 
solver in Section~\ref{sec:COMSOL} on a simple mesh structure.
Next, we apply our current-density-based and voltage-based 
approaches to the large public-domain IBM power grid benchmarks in Section~\ref{sec:IBM}. 
Finally, in Section~\ref{sec:OpeNPDN}, we perform an analysis on more modern 
nodes: commercial 12nm FinFET and 28nm FDSOI nodes, and an open-source 45nm 
technology, all based on Cu DD interconnects. We implement both our voltage-based 
and current-density-based analyses in Python3.6 and apply them to the benchmarks. 

Although our method can be applied to interconnect structures built using any materials, we use
modern Cu DD technologies in our evaluations. In particular, since 
the IBM benchmarks were originally designed for Al interconnects, a 
technology that is now obsolete, we use the topologies from this benchmark 
set and assume the interconnect to be built with Cu DD wires. In 
Cu DD interconnects, each layer can be treated separately due
to the presence of barrier/capping layers that prevent atomic flux from flowing
across layers through vias~\cite{Gambino18,Zhang10}.  The methods in this paper are applied to each
layer to find the steady-state stress, which is then used to predict
immortality.  This limits the size of the EM problem, since it must be solved
in a single layer at a time.  Moreover, since it is common to use a reserved
layer model where all wires in a layer are in the same direction~\cite{Jhaveri10}, effectively
this implies that each layer consists of a set of metal lines with a limited
number of nodes.  In such scenarios, the EM problem reduces to the analysis of
a large number of line/tree structures.  However, our
method is also exercised on the IBM benchmarks, which contain mesh structures within 
layers, allowing for full evaluation of the generality of our method.

\subsection{Comparison with COMSOL}
\label{sec:COMSOL}

\noindent
We show comparisons between our approach and numerical simulations using
COMSOL on Cu DD structures.  The material parameters, provided to COMSOL,
are~\cite{ala:05}:
$\rho =$ 2.25e-8$\Omega$m,
${\cal B} =$ 28GPa,
$\Omega =$ 1.18e-29m$^3$,
$D_0 =$ 1.3e-9m$^2$/s,
$E_a =$ 0.8eV,
$Z^* =$ 1,
$\sigma_{crit} = 41$MPa,
$T = 378$K.
Note that since $\beta= (Z^* e \rho)/\Omega$, the constant $\beta/\rho$ used in Eq.~\eqref{eq:sigmai_rewritten_voltage} of the voltage-based formulation is $(Z^* e)/\Omega$. 

COMSOL is limited to analyzing small structures, which is reflected 
the topologies shown in Fig.~\ref{fig:COMSOL_color_map}:
\begin{itemize}
\item
An interconnect tree with three segments
\item
A larger interconnect tree
\item
A simple mesh structure
\end{itemize}
The color maps in the figure show the spatial variation of steady-state stress
over each interconnect. The numbers next to each node represent the
values computed using our approach and by COMSOL. The
numbers match well; our approach is exact, and the small discrepancies are due
to numerical inaccuracies in COMSOL, e.g., due to discretization.

\subsection{Analysis on IBM power grid benchmarks}
\label{sec:IBM}

\noindent
The only widely used power grid benchmark suite is the set of IBM
benchmarks~\cite{IBMPDN_url}.  Each benchmark contains Vdd and Vss networks and
multiple voltage domains, and general tree/mesh structures in individual
layers. We use SPICE to obtain the branch current and node voltages. For these benchmarks, we use two approaches for stress computations (i)~a current-density-based traversal as explained in Section~\ref{sec:solution}, and (ii)~a voltage-based approach as explained in Section~\ref{sec:voltage_solution}. 

The IBM power grid benchmarks are available as SPICE netlists and do 
not specify widths and thicknesses of segments in the grid. 
Therefore, we back-calculate the product of the width and thickness 
(cross-sectional area) of each segment such that, in consistency with
Eq.~\eqref{eq:jL_IR}, the $jl/(IR)$  remains constant within
a metal layer, i.e., the cross-sectional area  of the segment is 
the reciprocal of the resistivity.

We implement a BFS traversal over these structures using Python3.6
and Deep Graph Library~\cite{karypis20} by modifying the message
passing functions. For both approaches, a single traversal is required
to find the connected components in the graph, on which the computations
are carried out. For the current-density-based formulation, a BFS traversal
is used to compute the stress value at every node, while the voltage-based 
formulation computes the stress at every nodes
using~\eqref{eq:sigmai_rewritten_voltage} without a traversal. 
We report the runtimes on a  2.2GHz Intel 
Xeon 
Silver 4114 CPU for both approaches.

\begin{figure}[htb]
\centering
\includegraphics[width=0.63\linewidth]{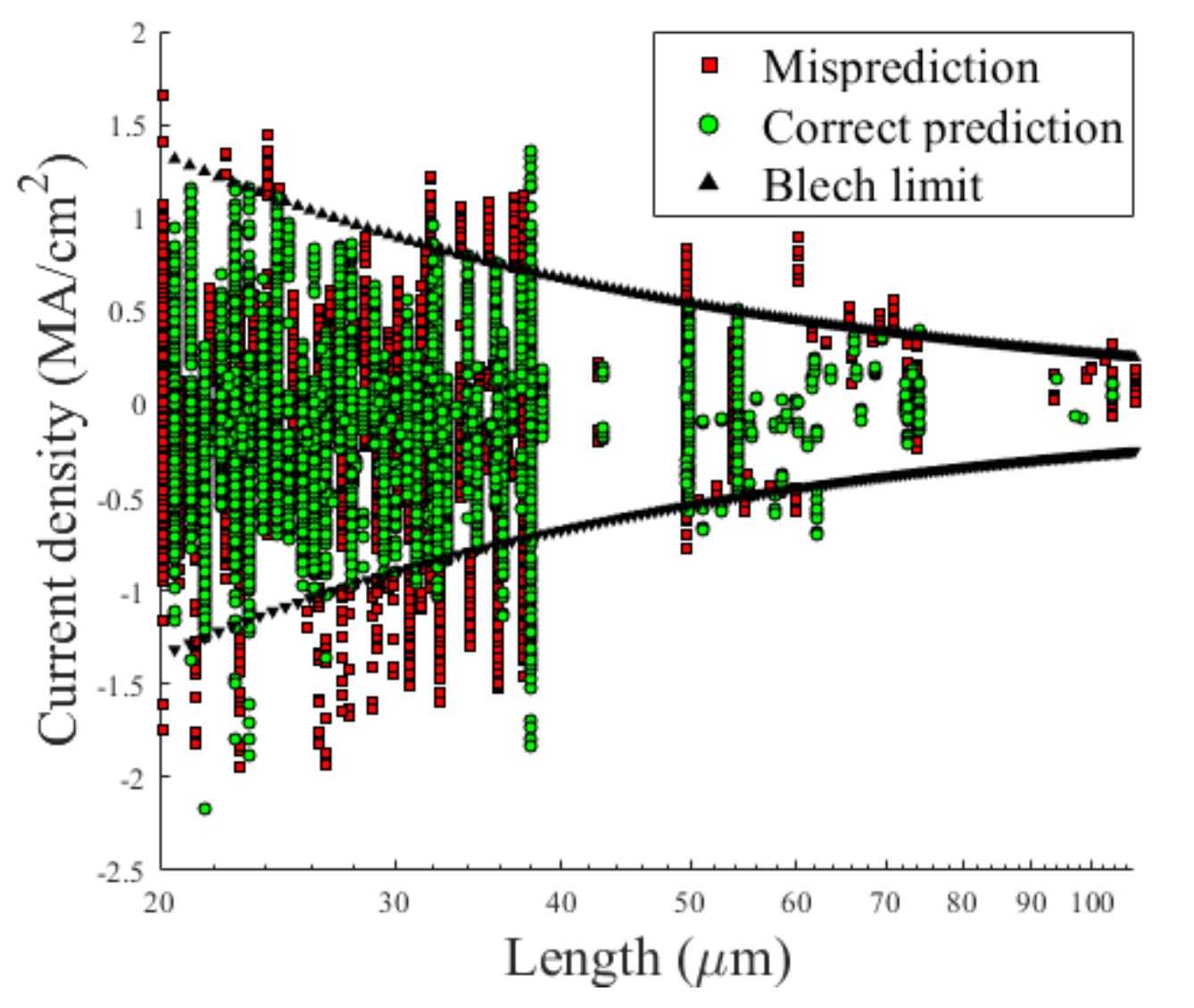}
\caption{Inaccuracy of the traditional Blech filter (ibmpg6).}
\label{fig:scatter-ibmpg6}
\end{figure}

The traditional Blech criterion is only accurate for a single-segment wire:
next, we evaluate its accuracy.  We consider our approach as the accurate
result since it is rigorously derived for multisegment structures by
generalizing the same physics-based modeling framework used by the Blech
criterion for one-segment wires, and it is validated on COMSOL.  Therefore a
positive identification of immortality implies that our method
finds the segment to be immortal; a negative identification implies mortality.
Fig.~\ref{fig:scatter-ibmpg6} plots the current density $j$ vs. the wire length
$l$ within the segments of the ibmpg6 benchmark\footnote{The figure shows the 1.6M edges of the ibmpg6 benchmarks. The scatter points of several segments may be hidden due to overlaps.}. The currents in the Vdd and
Vss lines may be either positive or negative, and their magnitude affects EM.
The black triangles show the contours of $jl = (jl)_{crit}$: when the magnitude
lies within this frontier for a segment of the grid, the traditional Blech
criterion~\eqref{eqn:Blech_criterion} would label the wire as immortal;
otherwise it is potentially mortal.  To help highlight erroneous predictions,
the figure shows green markers for correct predictions and red markers for
incorrect predictions. The Blech criterion shows significant inaccuracy 
on multisegment wires.

\begin{table}
\centering
\caption{Comparison of our approach against the traditional Blech filter on the IBM benchmarks (TP = true positive, TN = true negative, FP = false positive, FN = false negative.)}
\label{tbl:table-ibm}
\resizebox{\linewidth}{!}{%
\begin{tabular}{||c||r||r|r|r|r||r|r||} 
\hhline{|t:=:t:=:t:====:t:==|}
 & \multicolumn{1}{c||}{$|E|$} & \multicolumn{1}{c|}{TP} & \multicolumn{1}{c|}{TN} & \multicolumn{1}{c|}{{\bf FP}} & \multicolumn{1}{c||}{\bf{FN}} & \multicolumn{2}{c|}{Runtime (s)} \\ 
\cline{7-8}
 &     &        &           &              &                  & J-based  & V-based  \\
\hhline{|t:=:t:=:t:====:t:==|}
pg1 & 29750 & 7788 & 7432 & \textbf{9079} & \textbf{5451} & 6.8 & 4.0 \\ 
\hline
pg2 & 125668 & 44564 & 18943 & \textbf{45224} & \textbf{16937} & 17.6 & 9.5 \\ 
\hline
pg3 & 835071 & 481604 & 4328 & \textbf{346322} & \textbf{2817} & 119.4 & 80.6 \\ 
\hline
pg6 & 1648621 & 1173842 & 177 & \textbf{473122} & \textbf{1480} & 243.66 & 150.3 \\
\hhline{|b:=b:=:b:====:b:==|}
\end{tabular}
}
\end{table}

We compare the predictions of the traditional Blech criterion against the
ground truth, which corresponds to the provably correct analysis from our method. 
True predictions (true positive (TP) and true negative (TN)) correspond 
to correct predictions where the Blech criterion agrees with our accurate 
analysis; otherwise the predictions are false (false positive (FP) and 
false negative (FN)).  A positive prediction from the Blech criterion implies 
that a segment is immortal; a negative prediction indicates a mortal segment. The 
errors correspond to false negative predictions, where a truly immortal segment is deemed
potentially mortal by the traditional Blech criterion, and 
FPs, where a mortal segment is labeled as potentially immortal by
Blech. False positives cause failures to be overlooked, and false negatives may lead to
overdesign as EM-immortal wires are needlessly optimized.
Table~\ref{tbl:table-ibm} summarizes the results on IBM benchmarks\footnote{We do not show the numbers for pg4 and pg5 as we find all segments in these benchmarks to be immortal in our experiments.}. The table shows that:
\begin{itemize}
\item
the inaccuracies in the traditional Blech filter are not isolated but are seen across benchmarks.
\item
our method is scalable to large mesh sizes with low runtimes.
\end{itemize}
From the data, it is apparent that the traditional Blech criterion can provide
misleading results. The reasons for this are:
\begin{itemize}
\item
A high-$jl$ segment could be immortal if it has numerous downstream segments
with low $jl$, so that the total $jl$ sum may be low. For example, in Fig.~\ref{fig:twoseg}, if the current density $j_1 = 0$, then the segment acts as passive
reservoir, bringing down the stress in the right segment to be lower than
the case of an identical isolated segment carrying the same current, but with
a blocking boundary at $v_2$~\cite{Lin16}.  
\item
A low-$jl$ segment could be labeled immortal by the traditional criterion,
but it may be mortal due to a high stress at one node, caused by a high Blech
sum for downstream wire segments, which could raise the stress at the other
node.
\end{itemize}

\begin{figure}
\centering
\includegraphics[width=0.75\linewidth]{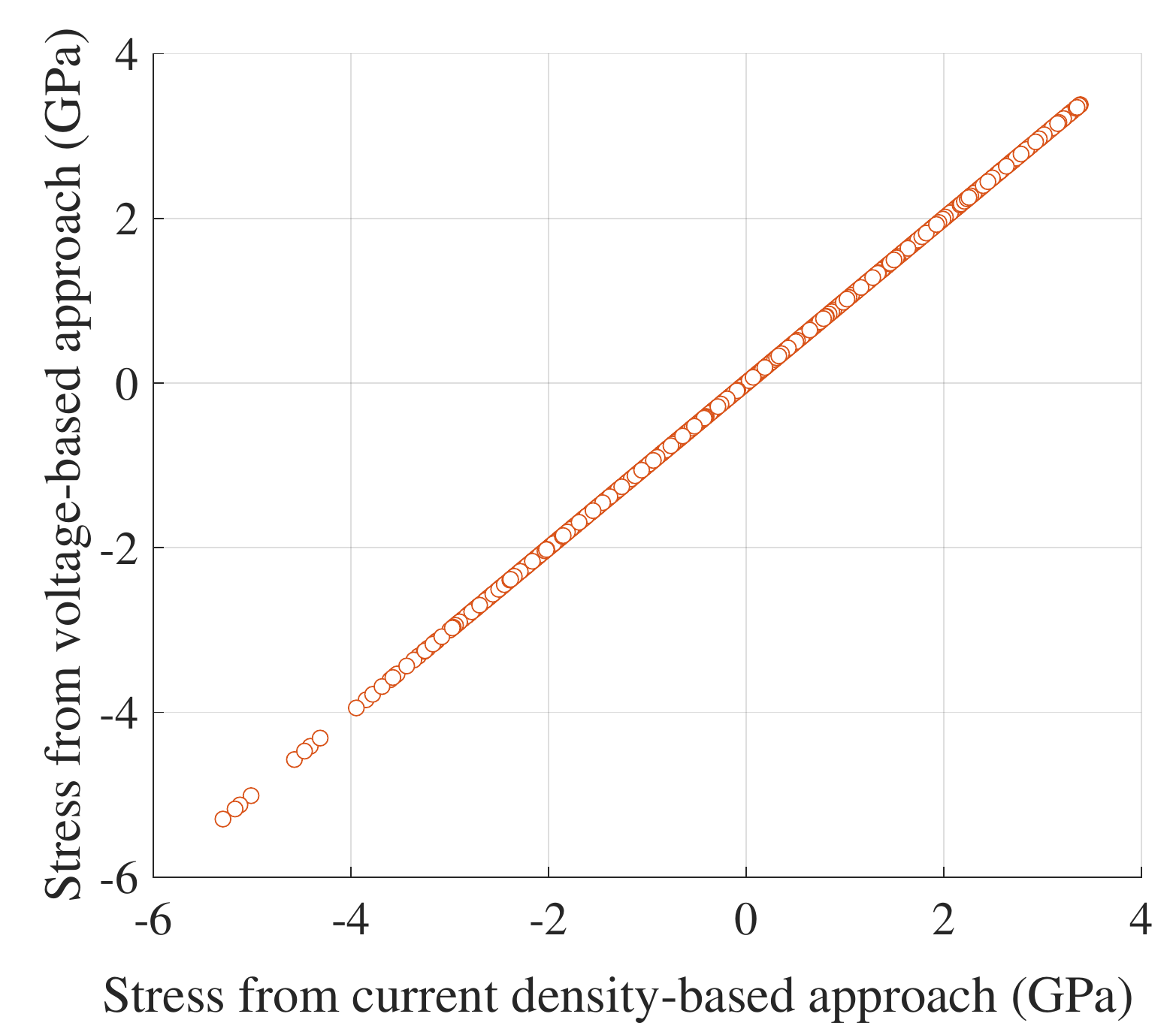}
\caption{Comparison of stress values from voltage-based and current-density-based approaches for ibmpg1.}
\label{fig:j-vs-v}
\end{figure}

We verify that our voltage-based and current-based formulations are identical by comparing the computed stress values for all segments in the IBM benchmarks. For example, Fig.~\ref{fig:j-vs-v} shows a scatter plot of the stress values for all segments in the ibmpg1 benchmark where the x-axis has the stress values computed by the current-density-based approach and y-axis shows the stress value of the segments computed by the voltage-based approach.  
The mean error is 18pPa and the maximum error is less than 60pPa. These negligibly small errors can be attributed to numerical precision issues.  

The TP, TN, FP, and FN values in Table~\ref{tbl:table-ibm} are identical between our equivalent voltage-based and current-density-based approaches. 
The table shows that the voltage-based approach is 1.5--1.9$\times$ faster than the current-density-based approach as it does not require traversals for stress computation.\footnote{We do not include the runtimes for parsing the benchmark files as modern design flows work with databases where power grid nets and wires can be queried in negligible time.}

\subsection{Analysis on OpenROAD power grids}
\label{sec:OpeNPDN}

\noindent
In this section, we show simulations based on power grids from circuits designed 
using a commercial 12nm FinFET technology, a 28nm technology, and an open-source 
45nm technology using Cu DD interconnects.  The
circuits are taken through synthesis, placement, and routing in these technology nodes
(some circuits are implemented in both nodes) using a standard design flow.
The power grid is synthesized using an open-source tool,
OpeNPDN~\cite{Chhabria20OpeNPDN} from OpenROAD.  The IR drop and currents are
computed using PDNSim~\cite{PDNSim}. Since the standard cell rows in the OpenROAD 
benchmarks have low utilizations, the current densities in the chip are low. 
Therefore, to evaluate our method, we scale the branch currents such that there 
are tens of mortal segments in each design.\footnote{In principle, the same effect would
be achieved with a sparser power grid, with larger IR drops, and therefore 
larger $jl$ values, i.e., larger Blech sums that translate into more EM failures.}

\begin{figure}[b]
\centering
\includegraphics[width=0.65\linewidth]{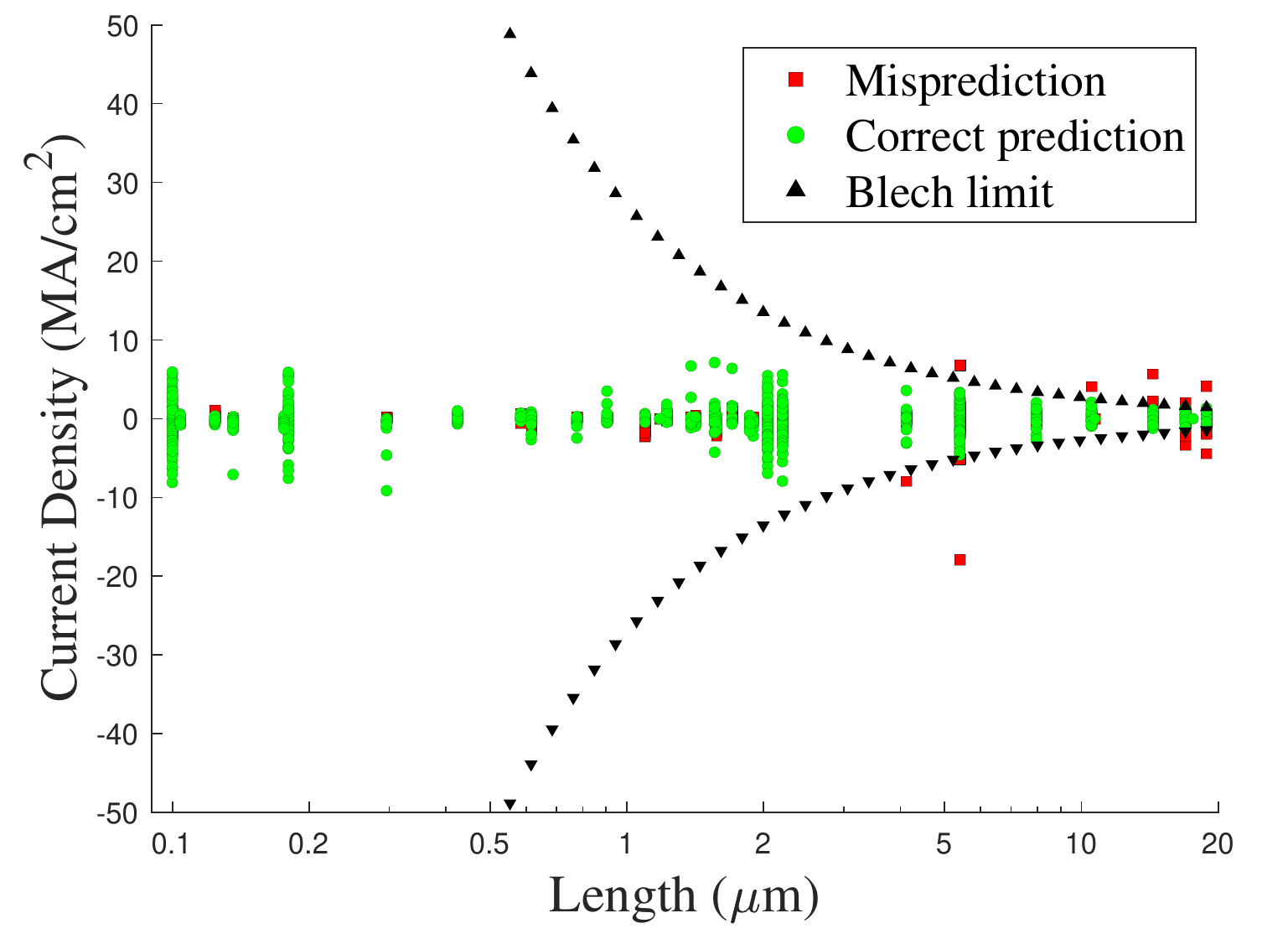}
\caption{Inaccuracy of traditional Blech filter (aes/12nm).}
\label{fig:scatter-OpeNPDN}
\end{figure}

Fig.~\ref{fig:scatter-OpeNPDN} shows a scatter plot that analyzes the
inaccuracy of the traditional Blech criterion on a Cu DD technology, using
$(jl)_{crit} = 0.27$A/$\mu$m, based on material parameters listed in
Section~\ref{sec:COMSOL}. Due to the regular structure of the power grid, 
many lines have the same length. As in the earlier case, it is easily seen that
the Blech criterion leads to numerous false positives and false negatives.
Results for more circuits are listed in Table~\ref{tbl:OpeNPDN} and show
similar trends. The number of mortal segments as per Blech criterion, i.e.,
true negatives and false negatives is small across all benchmarks and 
technologies. This is attributed to the fact that $jl$ values are small 
as compared to ibmpg benchmarks. However, there are significant numbers of 
false positives across all benchmarks, which indicates the inaccuracy of
the traditional Blech filter, when misused for multisegment wires. For these
testcases, the voltage-based solution is 1.7--2.7$\times$ faster than
the current-density-based solution.

\begin{table}
\centering
\caption{Comparison of our approach against the traditional Blech filter on various technology with Cu interconnects.}
\label{tbl:OpeNPDN}
\resizebox{\linewidth}{!}{%
\begin{tabular}{||c||l||r||r|r|r|r||r|r||} 
\hhline{|t:=:t:=:t:=:t:====:t:==:t|}
\multirow{2}{*}{} & Circuit & \multicolumn{1}{c||}{$|E|$ } & \multicolumn{1}{c|}{TP} & \multicolumn{1}{c|}{TN} & \multicolumn{1}{c|}{\bf FP} & \multicolumn{1}{c||}{\bf FN} & \multicolumn{2}{c||}{Runtimes (s)} \\ 
\hhline{||~||~||~||~|~|~|~|:==:|}
 &  & \multicolumn{1}{c||}{} & \multicolumn{1}{c|}{} & \multicolumn{1}{c|}{} & \multicolumn{1}{c|}{} & \multicolumn{1}{c||}{} & \multicolumn{1}{c|}{J-based} & \multicolumn{1}{c||}{V-based} \\  
\hhline{|:=::=::=::====::==:|}
\multirow{4}{*}{12nm} 
 & gcd & 4,121   & 2,177 & 94 & {\bf 1,821} & {\bf 29} & 0.9 & 0.5 \\ 
\hhline{||~|:=::=::====::==:|}
 & jpeg & 83,743 & 50,436 & 5 & {\bf 33,250} & {\bf 52} & 10.6 & 6.2 \\ 
\hhline{||~|:=::=::====::==:|}
 & dynamic\_node & 150,768 & 79,990 & 0 & {\bf 70,757} & {\bf 21} & 17.9 & 9.5 \\ 
\hhline{||~|:=::=::====::==:|}
 & aes & 194,485 & 84,132 & 0 & {\bf 110,330} & {\bf 23} & 23.3 & 12.3 \\ 
\hhline{|:=::=::=::====::==:|}
\multirow{2}{*}{28nm}
 & gcd & 678 & 400 & 70 & {\bf 158} & {\bf 50} & 0.4 & 0.2 \\ 
\hhline{||~|:=::=::====::==:|}
 & aes & 11,361 & 4,946 & 62 & {\bf 5,862} & {\bf 491} & 2.9 & 1.6 \\ 
\hhline{|:=::=::=::====::==:|}
\multirow{4}{*}{45nm} 
& dynamic\_node & 6,610 & 4,943 & 0 & {\bf 1,641} & {\bf 26} & 1.2 & 0.6 \\ 
\hhline{||~|:=::=::====::==:|}
 & aes & 7,996 & 5,562 & 2 & {\bf 2,393} & {\bf 39} & 1.9 & 0.7 \\ 
\hhline{||~|:=::=::====::==:|}
 & ibex & 12,723 & 9,273 & 0 & {\bf 3,438} & {\bf 12} & 1.8 & 0.8 \\ 
\hhline{||~|:=::=::====::==:|}
 & swerv & 61,935 & 43,122 & 0 & {\bf 18,810} & {\bf 3} & 7.4 & 4.0 \\
\hhline{|b:=:b:=:b:=:b:====:b:==:b|}
\end{tabular}
}
\end{table}

\begin{figure}[tb]
\centering
\includegraphics[width=0.65\linewidth]{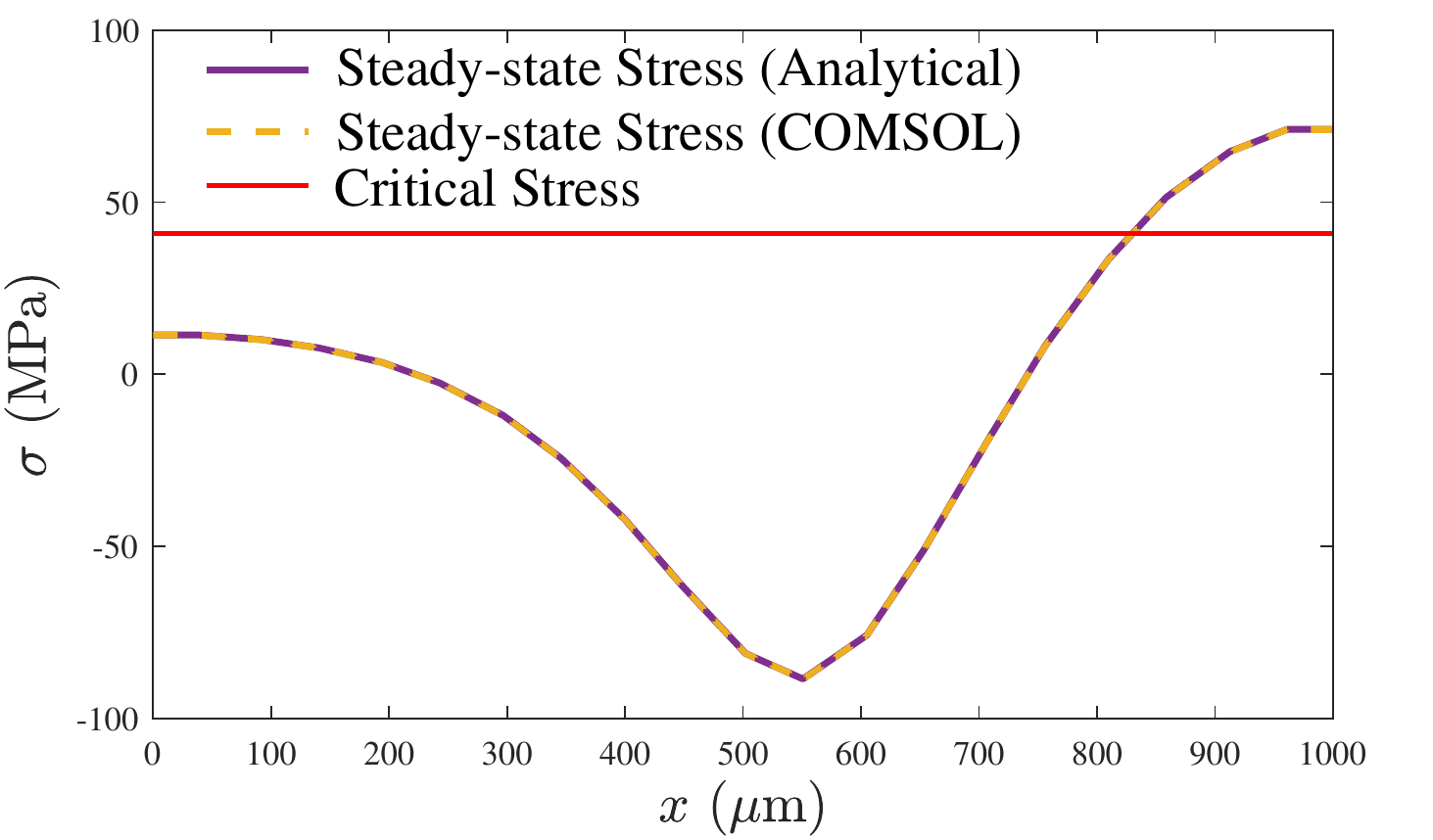}
\caption{Stress profile of a 185-segment line of length 999$\mu$m in aes synthesized in a commercial 12nm FinFET technology. The segment current density values range from 81.5MA/m$^2$ to 1.92GA/m$^2$.}
\label{fig:gf12-jpeg-single-line}
\end{figure}

Next, we show the stress profile in a single power stripe
of length 999$\mu$m from the aes benchmark, synthesized in a commercial 12nm technology. 
The line consists of 185 segments, each carrying different currents. The stress profile is
shown in Fig.~\ref{fig:gf12-jpeg-single-line}: due to the large number of segments in
the line, the profile may appear nonlinear at this resolution, but the steady-state 
stress in each segment is linear along its length, and the stresses obey continuity at
the segment boundaries, as specified by the boundary conditions. The figure shows
excellent agreement between our analysis and a COMSOL simulation. However, COMSOL 
must solve the transient stress problem for a long period before the steady state 
is achieved, while our approach provides the solution in milliseconds.

\section{Conclusion}
\label{sec:conclusion}

\noindent
This work proposes a theoretically justified method for checking immortality in a general tree
or mesh interconnect. The theoretical basis for the method is presented, and two versions --
a current-density-based method and a voltage-based approach -- are presented. Although not elaborated
upon in this work, the voltage-based formulation is potentially useful for power grid optimization, 
since it translates EM constraints into IR constraints, enabling a unified formulation that optimizes
a power grid for both IR drop and EM, while operating purely in the realm of voltages (i.e.,
translating stress variables to voltage variables).  Both the current-density-based and voltage-based
approaches have linear time complexity, the latter is, on average, 1.9$\times$ faster than the former. 
The results are validated  against COMSOL and it is shown that the methods are fast and 
scalable to large power grids.

\bibliographystyle{misc/ieeetr2}
\bibliography{bib/main,bib/main3}
\end{document}